\documentclass[[11 pt, a4paper,table]{article}
\usepackage[latin9]{inputenc}
\usepackage{amsmath}
\usepackage{mathtools}
\usepackage[unicode=true,pdfusetitle,
 bookmarks=true,bookmarksnumbered=false,bookmarksopen=false,
 breaklinks=false,pdfborder={0 0 1},backref=false,colorlinks=false]
 {hyperref}

\makeatletter

\DeclareTextSymbolDefault{\textquotedbl}{T1}

\@ifundefined{date}{}{\date{}}
\usepackage{tikz}
 \usepackage{stmaryrd}
  \usepackage{stackrel}

\usepackage{subfig}
\usepackage{rotating}
\usepackage{tabularx}
\usepackage{graphicx}
\usepackage{afterpage}
\usepackage{array}
\usepackage{multirow}

\usepackage{latexsym}
\newcommand{\be}{\begin{equation}}
\newcommand{\ee}{\end{equation}}
\newcommand{\bea}{\begin{eqnarray}}
\newcommand{\eea}{\end{eqnarray}}
\newcommand{\beas}{\begin{eqnarray*}}
\newcommand{\eeas}{\end{eqnarray*}}

\def\tr{{\rm Tr}}

\def\XXint#1#2#3{{\setbox0=\hbox{$#1{#2#3}{\int}$ }
\vcenter{\hbox{$#2#3$ }}\kern-.5\wd0}}

\makeatother

\begin{document}
\title{Large $N$ Master Field  Optimization: the Quantum Mechanics of two Yang-Mills coupled Matrices}
\author{Kagiso Mathaba\thanks{Email: kgs.mathaba@gmail.com}, 
Mbavhalelo Mulokwe\thanks{Email: mbavhalelo.mulokwe@wits.ac.za} 
and Jo\~ao P. Rodrigues\thanks{Email: joao.rodrigues@wits.ac.za} 
\\
 \\
 National Institute for Theoretical and Computational Sciences\\
 School of Physics and Mandelstam Institute for Theoretical Physics
\\
 University of the Witwatersrand, Johannesburg\\
 Wits 2050, South Africa \\
 }
\maketitle
\begin{abstract}
We study the large $N$ dynamics of two massless Yang-Mills coupled matrix quantum mechanics, by minimization of a loop truncated Jevicki-Sakita effective collective field Hamiltonian. The loop space constraints are handled by the use of master variables. The method is successfully applied directly in the massless limit for a range of  values of the Yang-Mills coupling constant, and the scaling behaviour of different physical quantities derived from their dimensions are obtained with a high level of precision. We consider both planar properties of the theory, such as the large $N$ ground state energy and multi-matrix correlator expectation values, and also the spectrum of the theory. For the spectrum, we establish that the $U(N)$ traced fundamental constituents remain massless and decoupled from other states, and that bound states develop well defined mass gaps, with the mass of the two degenerate lowest lying bound states being determined with a particularly high degree of accuracy. In order to confirm, numerically, the physical interpretation of the spectrum properties of the $U(N)$ traced constituents, we add masses to the system and show that, indeed, the $U(N)$ traced fundamental constituents retain their "bare masses". For this system, we draw comparisons with planar results available in the literature. 
 \end{abstract}

\section{Introduction}
Multi-matrix systems \footnote{We have in mind the path integral or the quantum mechanics of a finite number of matrices.} are important in many different physical settings, not only as of finite sized matrices, but particularly in their large $N$ limit \cite{tHooft:1973alw}. They are expected to provide a reduced ansatz for large $N$ gauge theories, both in the path integral 
\cite{Eguchi:1982nm, Bhanot:1982sh, Parisi:1982gp, Gross:1982at, Das:1982ux} and Hamiltonian formulations \cite{Neuberger:1982ne, Kitazawa:1982wn}, and both for unitary as well as hermitian matrices. Their interpretation as $D0$ branes \cite{Polchinski:1995mt} has led to the suggestion that they may provide a definition of M theory\cite{Banks:1996vh}, also argued to be valid for the path integral \cite{Ishibashi:1996xs}.  The $AdS/CFT$ correspondence \cite{Maldacena:1997re, Gubser:1998bc, Witten:1998qj} has highlighted the importance of $\mathcal{N}=4 $ SYM theory, with its bosonic adjoint scalar sector, and ensuing integrability properties \cite{Beisert:2010jr} and Hamiltonian reductions \cite{Berenstein:2002jq},  \cite{deMelloKoch:2002nq, Beisert:2002ff, Kim:2003rza}. They are used in the study of black holes \cite{Kazakov:2000pm, Cotler:2016fpe, Maldacena:2023acv}\footnote{There is a vast literature on matrix models; we have tried to highlight only some key developments in their application, with emphasis on YM coupled systems.}.

We study in this communication the large $N$ properties of the Hamiltonian of two massless matrices interacting via a Yang-Mills potential. We are able to study the system directly in the large $N$ limit and directly in the massless limit. In general \cite{Morita:2020liy, Pateloudis:2022ijr}, properties of the massless system are obtained by extrapolation of a system with a finite mass parameter to zero. In our case, being able to work directly with the massless system, we are able to obtain the asymptotic scaling behaviour of physical quantities and correlators, and determine their parameters, with a high degree of accuracy. 

In addition to its interpretation as a reduced model, $D0$ or bosonic scalar subsector of  $\mathcal{N}=4 $ SYM, this Hamiltonian has associated with it a number of puzzles, mainly around the properties of its spectrum and of the presence or not of a mass gap \cite{Simon:1983jy, deWit:1988wri, Froehlich:1997df}.  Within the context of the $1/N$ expansion, we provide a definite answer to this question in this article. 

Our approach is based on the collective field theory Hamiltonian of Jevicki and Sakita \cite{Jevicki:1979mb}. This Hamiltonian is an exact re-writing of a given theory in terms of its (gauge) invariant variables. The large $N$ (planar) background is then obtained semiclassically as the minimum of an effective potential $V_{eff}$ and, when expanded about this large $N$ background, the collective field theory Hamiltonian generates $1/N$ corrections systematically \footnote{For a single matrix based example, see for instance  \cite{Das:1990kaa},    \cite{Demeterfi:1991cw}. Also, although we consider matrix valued systems in this communication, the same is true of vector valued field theories.}.

The idea behind the method is to implement a change of variables from the original variables of the theory, generically denoted by $X_\mathcal{A}$, to the invariant set of operators (the collective fields), generically denoted by $\phi(C)$, and to require explicit hermiticity of the collective field Hamiltonian. This change of variable is accompanied by a Jacobian $J$. In general $J$ is not known explicitly, but it satisfies the following equation
\begin{equation*}
\sum_{C'}\frac{\partial \ln J}{\partial \phi^{\dagger}(C')} \Omega(C',C) = w(C) -  \sum_{C'}\frac{\partial \Omega(C',C)}{\partial \phi^{\dagger}(C')}.\end{equation*}
This is sufficient to obtain explicitly the collective field Hamiltonian in terms of $\phi(C)$ and its canonical conjugate $\pi(C)$. In general, 
\begin{equation*}
\Omega(C,C') =\sum_\mathcal{A} \frac{\partial \phi^{\dagger}(C) }{\partial X^{\dagger}_\mathcal{A}} \frac{\partial \phi(C') }{\partial X_\mathcal{A}} , \hspace{8pt}
w(C) =  \sum_A \frac{\partial^2 \phi(C) }{\partial X^{\dagger}_\mathcal{A} \partial X_\mathcal{A}} \, . \end{equation*}
$\Omega(C,C')$ joins two loops into a sum of single loops\footnote{The terminology "loop" is inherited from gauge theories, where the gauge invariant variables are Wilson loops. But in this communication they refer to operators that are invariant under the gauge symmetries allowed by the system under consideration.}, and $w(C)$ splits a given loop into a sum of two (in general smaller) loops. Details of their form will be given in the following.

The collective field Hamiltonian $H_{col}$ is ideally suited to a numerical approach based on minimisation of the effective potential $V_{eff}$, in a truncated loop space $H_{col} \to H^{trunc}_{col}$ that can be systematically increased and ascertained for convergence and accuracy.  Indeed, already some time ago \cite{Jevicki:1982jj, Jevicki:1983wu}, this approach was successfully implemented in the context of $2+1$ lattice gauge theories with Wilson's one-plaquette action \cite{Wilson:1974sk}. 
Systems of unitary matrices have a phase transition between a strong and weak phase, and it was established in \cite{Jevicki:1982jj, Jevicki:1983wu} that in the weak coupling phase the minimization has to be accompanied by a constraint:
\begin{equation}
\begin{cases}
      \text{Minimize}  \hspace{5pt} V_{eff}^{trunc}  ,\\ \Omega(C,C') \succeq 0. 
     \end{cases}    
 \end{equation}
In other words, the large $N$ expectation values of the loop variables $\phi(C)$ must satisfy the constraint that the matrix $ \Omega(C,C')$ is semi-positive definite, with a number of eigenvalues saturating to zero in the weak coupling regime. This was shown to also be the case when considering loop equations  \cite{Rodrigues:1985aq}.

This constraint is not difficult to understand: the large $N$ limit of the single unitary matrix integral has a well known third order phase transition \cite{Gross:1980he}, described in terms of the density of its (phases of) eigenvalues $\rho(\theta)$ as:
\begin{equation}
\begin{cases}
\rho(\theta)=  \frac{1}{2\pi} (1 + \frac{2}{\lambda} \cos \theta), \,\hspace{80pt} -\pi \le \theta \le \pi  &\text{for} \,\, \lambda \ge 2 , \\
\left\{\begin{array}{lr}
\rho(\theta)=\frac{2}{\pi \lambda} \cos \frac{\theta}{2} \sqrt{\frac{\lambda}{2} - \sin^2 \frac{\theta}{2}},   & |\theta| < 2 \sin^{-1} \sqrt{\frac{\lambda}{2}}\\
\rho(\theta)=0, & 2 \sin^{-1} \sqrt{\frac{\lambda}{2}} \le |\theta| \le \pi \end{array}\right\}    &\text{for} \,\,  \lambda \le 2  .\end{cases}
\end{equation}
In the strong coupling regime, the density of eigenvalues is periodic with period $2\pi$. For weak coupling, the density of eigenvalues develops finite support within the interval $[-\pi,\pi]$, and $\rho(\theta) = 0$ outside this finite support. 

A similar phase transition and pattern in the density of eigenvalues is present in the large $N$ limit of the quantum mechanics of single unitary matrix systems  \cite{Jevicki:1980zq, Wadia:1980cp, Rodrigues:1981sd, Rodrigues:1982qr}. Systems of hermitian matrices have only one single (weak) phase, so ensuring that $\phi(x)=0$ outside their finite support in order that the density of states remains non-negative is of paramount importance. 

For a single hermitian $N \times N$ matrix $M$, with invariants $\phi_{k} = \tr(e^{-ikM})= \sum_{i=1}^N e^{-ik\lambda_i}$, the density of eigenvalues is simply its Fourier transform, $\phi(x) = \int dk/2\pi \, e^{ikx} \phi_k = \sum_{i=1}^N \delta(x-\lambda_i)$. Then:
\begin{equation*}
\Omega(k,k') = k k' \phi_{k'-k},  \hspace{10pt} \Omega(x,y) = \partial_x \partial_y \left( \phi(x) \delta(x-y) \right).
\end{equation*}
So, in this simple case, $\Omega(x,y)$ is seen to have zero eigenvalues when the density matrix $<x|\hat{\phi}|y>= \phi(x) \delta(x-y)$ has zero eigenvalues, or when $\phi(x) = 0$.  For single matrix systems then, this constraint on $\Omega$ is easily related to the requirement that the density is non-negative.

In the case of more complex multi-matrix systems, and in the context of the collective field theory approach, the required constraint is then that the loop space matrix 
\begin{equation*}
\Omega(C,C') =\sum_\mathcal{A} \frac{\partial \phi^{\dagger}(C) }{\partial X^{\dagger}_\mathcal{A}} \frac{\partial \phi(C') }{\partial X_\mathcal{A}} = \sum_{C"} y(C,C',C") \phi(C")
\end{equation*}
is semi-positive definite. This follows from the definition of $\Omega(C,C')$, on the left hand side of the above equation, but no longer apparent once $\Omega(C,C')$ is expressed in terms of loop variables, as on the right hand side of the above equation. The semi-positiveness condition results in a number of inequalities amongst the loop variables. These are the non-linear constraints that large $N$ expectation values of invariant operators have to satisfy, some of which are saturated. For instance, if the $X_\mathcal{A}$'s are hermitian  matrices, the operation
$\frac{\partial}{\partial X_\mathcal{A}}$ acting on a single trace loop of a product of these matrices removes $X_\mathcal{A}$  everywhere where it is present along the loop and "opens up an open string". $\Omega(C,C')$ can then be thought of as a matrix of (linear combinations of) open string inner products, linking to the interpretation  as suggested in  \cite{Anderson:2016rcw}. \footnote{For a unitary matrices, $\partial /\partial X_{ab} \to U_{bc} \partial / \partial U_{ab} \left( = - U^{\dagger}_{ca}/ \partial U^{\dagger}_{cb} \right) $. }

Interest in the development of numerical techniques applicable directly to the large $N$ limit properties of multi matrix systems has been rekindled recently \cite{Anderson:2016rcw, Lin:2020mme, Han:2020bkb, Kazakov:2021lel, Koch:2021yeb}, 
and the importance the loop space constraints re-discovered \cite{Anderson:2016rcw, Lin:2020mme, Han:2020bkb, Kazakov:2021lel}, in a program generically referred to as numerical bootstrap. In  \cite{Anderson:2016rcw, Kazakov:2021lel, Kazakov:2022xuh} the constrained optimisation problem uses semidefinite programming in solving loop equations directly in loop space. In \cite{Lin:2020mme}, the loop equations are iterated and a set of small loops adjusted to satisfy the constraints, and this method is generalized to matrix quantum mechanics in \cite{Han:2020bkb}, where some exact bounds are also obtained. 

In this article we carry out a numerical study of the large $N$ limit of the quantum mechanics of two hermitian matrices coupled by a Yang-Mills potential using a truncated collective field Hamiltonian. We are able to carry out this study directly in the massless limit, with the different loop quantities considered exhibiting the expected scaling properties with an extremely high degree of accuracy. As indicated earlier, since the collective hamiltonian is an exact loop space re-writing of the large $N$ theory, it has a systematic $1/N$ expansion. In particular then, and as a quantum mechanical system, we are able to study its mass spectrum in addition to the planar properties of the theory.  The study of the spectrum and ensuing absence or presence of mass gaps is of particular importance to matrix gauge theories. We believe that currently, this is the only method able to provide information about the large $N$ spectrum of the theory. 

The issue of constraints is addressed by the use of "master variables"  \cite{Jevicki:1983wu, Koch:2021yeb,Jevicki:1983hb}. These are variables that satisfy the constraints explicitly, of which the original variables are an example. In addition to satisfying the constraints in the planar limit, they can be used to set up the spectrum equations of the theory \cite{Jevicki:1983hb}. For two matrix systems, we keep one of the matrices diagonal and the other as an arbitrary $N \times N$ hermitian, so that there are $N(N+1)$ master variables. 

The fact that another parameter $N$ has been introduced in addition to the truncation parameter may seem undesirable. But the extensive study of two matrix hamiltonians with cubic, quartic and mixed quadratic potential carried in \cite{Koch:2021yeb} as well as further evidence provided in this article establish the stability and accuracy of this approach\footnote{It also opens the possibility of direct studies of stringy (gravity and brane) phenomena emerging from properties of single trace operators of length $l$ related to different powers of $N$, as embedded in the AdS/CFT correspondence \cite{Maldacena:1997re,Gubser:1998bc,Witten:1998qj},\cite{Berenstein:2002jq},\cite{Lin:2004nb}. This is beyond the scope of this communication.   }.

Loop equations of the $d=0$ integral correspond to the statement of minimization of the large $N$ loop collective effective potential of a suitably defined quantum mechanical Hamiltonian, namely the Fokker-Planck Hamiltonian  \cite{Rodrigues:1985aq} \footnote{This effective potential also gives the leading Large $N$ loop configuration of the bosonic sector of supersymmetric multi matrix Marinari-Parisi \cite{Marinari:1990jc} type models \cite{Jevicki:1991yk},\cite{Rodrigues:1992by} .}. This is the best way to understand that constraints, (which in the collective field approach are encapsulated in the statement that $\Omega(C,C') \succeq 0$) arise even in the context of loop equations \cite{Rodrigues:1985aq}, and as argued differently and more recently in \cite{Anderson:2016rcw}. 
An extensive study of two matrix loop equations with cubic, quartic and mixed quadratic potential was carried in \cite{Koch:2021yeb}, which were shown to be satisfied to a very high level of accuracy. In this article, however, we are interested in the matrix quantum mechanics Hamiltonian of two Yang-Mills coupled matrices.   

This article is organized as follows: after the current Introduction, Section $2$ describes the method, the loop truncation and the use of master variables in dealing with the loop space constraints. Their use in obtaining both the large $N$ loop background, and concomitant planar quantities, as well as the $1/N$ spectrum, is described. In Section $3$ we apply the method to the large $N$ limit of the quantum mechanics of two Yang-Mills coupled, massless matrices. We find that the method displays perfect stable convergence directly in this massless limit. Moreover, physical quantities exhibit the scaling behaviour determined by their dimensions with a very high level of accuracy, both for planar quantities and for the $1/N$ spectrum. In other words, the loop truncated collective field Hamiltonian is entirely consistent with the scaling properties of the full massless theory.  Only correlators of zero charge of the $SO(2)\simeq U(1)$ symmetry of the system are shown to develop non-zero planar expectation values, and we assign charges to spectrum states.   
For the spectrum, we argue that the two lowest states, the $U(N)$ traced constituent single particle states \footnote{These are absent for $SU(N)$.}, are numerically massless, and associate them with the non-interacting $U(1) \times U(1)$ subgroup of the system. Higher energy (bound) states develop well defined mass-gaps. In order to confirm our physical interpretation of the two lowest degenerate states as massless states, we add masses to the system in Section $4$, where it is verified indeed that their "bare mass" is not corrected. Mass corrected fits to the planar ground state energy, to the first few bound states and to planar correlators are presented, some of which are compared with (the few results available in) the literature. As for the massless case, it is confirmed that only zero charge correlators develop non-zero expectation values, and charges to spectrum states are assigned.
Section $5$ is left for a conclusion and future outlook. The Appendix discusses estimates of errors in our numerical method. 

\section{Method, loop truncation and master variables}

\hspace{10pt} 
We consider the quantum mechanics of two $N \times N$ hermitian matrices $X_A, \, A=1,2$, interacting via a Yang-Mills potential:
\begin{equation*}
\hat{H} = \frac{1}{2}  \sum_{A=1}^{2} \tr{P_A^2} +  \frac{m^2}{2} \sum_{A=1}^{2} \tr{X_A^2}  - \frac{g_{YM}^2}{N} \tr [X_1,X_2]^2 =  \frac{1}{2}  \sum_{A=1}^{2} \tr{P_A^2} + \tr (V(X_A)).\end{equation*}
$P_A$ is canonical conjugate to $X_A$, and $m$ is a mass. The massless case $m=0$ will be considered first in Section 3 and the massive case in Section 4. Note that in terms of our conventions,  't Hooft's coupling $\lambda$ is $\lambda= g_{YM}^2$.

The $U(N)$ invariant loops are single traces of products of the matrices $X_A$, up to cyclic permutations:
\begin{equation*}
\phi(C) = \tr (... X_1^{m_1} X_2^{m_2} X_1^{n_1} X_2^{n_2}...)\, .
\end{equation*}
For instance, with two matrices one has $[1\, 1]= \tr(X_1^2)\, , [1 \,2]=\tr(X_1 X_2)\, , [2\, 2]= \tr(X_2^2)$, with three matrices $[1 \,1 \,1] = \tr(X_1^3)\, , [1\, 1 \,2] = \tr(X_1^2 X_2)\, , [1\, 2 \,2] = \tr(X_1 X_2^2)\, , [2\, 2 \,2] = \tr(X_2^3)$, etc., with an obvious notation. We will continue to refer to the invariant variables as  ``loops", for the historical reasons explained in the introduction.

The collective field Hamiltonian \cite{Jevicki:1979mb} in terms of the invariant loops $\phi(C)$ takes the form 
\begin{equation*}
H'_{col} = \frac{1}{2} \sum_{C,C'}  \pi^{\dagger}(C) \Omega(C,C') \pi(C') + \frac{1}{8} \sum_{C,C'} w(C) \Omega^{-1}(C,C') w^{\dagger}(C') + V(\phi) + \Delta H' ,
\end{equation*}
where $\pi(C)$ is the canonical conjugate to $\phi(C)$, and  
\begin{eqnarray*}
\Omega(C,C') &=&\sum_{A=1}^2 \tr \left( \frac{\partial \phi^{\dagger}(C) }{\partial X^{\dagger}_A} \frac{\partial \phi(C') }{\partial X_A} \right) = \sum_{C"} y(C,C',C'') \phi(C'') \\
w(C) &=& \sum_{A=1}^2 \tr \left(  \frac{\partial^2 \phi(C) }{\partial X^{\dagger}_A\partial X_A} \right)  = \sum_{C',C"} z(C,C',C'') \phi(C')  \phi(C'')  .  \end{eqnarray*}
If $C$ has length (number of matrices in the loop) $l(C)$ and $C'$ has length $l(C')$,  $\Omega(C,C')$ joins them into a number of loops of length $l(C)+l(C')-2$. $w(C)$ splits the loop $C$ of length $l(C)$ into sets of two loops $C'$ and $C''$ with total lengths $l(C)-2$.
$\Delta H'$ contains subleading counterterms that need not be considered for the large N background and the spectrum.

We then consider 
\begin{eqnarray*}
H_{col} &=& \frac{1}{2} \sum_{C,C'} \pi^{\dagger}(C) \Omega(C,C') \pi(C') + V_{eff} (\phi) \, , \\
V_{eff} (\phi) &\equiv& \frac{1}{8} \sum_{C,C'} w(C) \Omega^{-1}(C,C') w^{\dagger}(C') + V(\phi) .
\end{eqnarray*}

In order to exhibit explicitly the large $N$ dependence, we let 
\begin{equation*}
\phi(C) \to \frac{\phi(C)}{N^{\frac{l(C)}{2}+1}} = \frac{\tr (... X_1^{m_1} X_2^{m_2} X_1^{n_1} X_2^{n_2}...)}{N^{\frac{l(C)}{2}+1}} , \hspace{6pt} \pi(C) \to N^{\frac{l(C)}{2}+1} \pi(C) \end{equation*}
and obtain
\begin{align}
H_{col} &= \frac{1}{2N^2} \sum_{C,C'} \pi^{\dagger}(C) \Omega(C,C') \pi(C') + N^2 V_{eff} (\phi) \, , \label{Hcoll}\\
V_{eff} (\phi) &\equiv \frac{1}{8} \sum_{C,C'} w(C) \Omega^{-1}(C,C') w^{\dagger}(C') + V(\phi) . \label{Veff}
\end{align} 

It follows that the large $N$ background is the minimum of $V_{eff}$ subject to the constraint that $\Omega(C,C')$ is semi-positive definite.\footnote{The discussion next in this section follows closely that of \cite{Koch:2021yeb}, which is based on \cite{Jevicki:1982jj,Jevicki:1983wu,Jevicki:1983hb}.}
\subsection{Truncation of loop space}
For a given $l$ ($l \ge 4$), $\Omega$ is truncated to be a $N_{\Omega} \times N_{\Omega}$ matrix, where $N_{\Omega}$ is the number of loops of length $l$ or less. $\Omega$ itself, however, depends on loops with lengths up to $l_{\rm max}=2 l -2$. If $N_{\rm loops}$ is the number of loops with length $l_{\rm max}$ or less, then it is seen that $V_{eff}$  in (\ref{Veff}) is a function of $N_{\rm loops}$:
\begin{equation*}
V^{trunc}_{eff} (\phi(C), C=1,...,N_{\rm loops}) = \frac{1}{8} \sum_{C,C' = 1}^{N_{\Omega}} w(C) \Omega^{-1}(C,C') w^{\dagger}(C') + V(\phi) \end{equation*}
These parameters are listed in the following table for different levels of truncation. 
\begin{table}[h]
\begin{center}
\begin{tabular}{||c|c|c||} 
\hline
$l_{\rm max}$& $N_\Omega$  & $N_{\rm loops}$ \\ [0.5ex] 
\hline\hline
4 & 9 & 15\\ 
\hline
6 & 15 &37\\ 
\hline
8 & 23 &93\\ 
\hline
10 & 37 &261\\ 
\hline
12 & 57 &801\\
\hline
14 & 93 &2615\\
\hline
16 & 153 &8923\\
\hline
18 & 261 &31237\\ [1ex] 
 \hline
\end{tabular}
\caption{Truncating loop space}
\label{table:1}
\end{center}
\end{table}

\subsection{Master fields and planar limit.}
In order to minimize $V^{trunc}_{eff}$ subject to the constraint $ \Omega(C,C') \succeq 0$, we introduce master variables $\phi_{\alpha}$ that explicitly satisfy the constraint:
\begin{equation*}
\Omega(C,C') =\sum_{\alpha} \frac{\partial \phi^{\dagger}(C) }{\partial \phi_{\alpha}} \frac{\partial \phi(C') }{\partial \phi_{\alpha}}\succeq 0\end{equation*}
Specifically, we choose $X_1$ to be diagonal and $X_2$ an arbitrary $N \times N$ hermitian matrix. The master field then has $N^2+N$ real components $\phi_{\alpha},\, \alpha=1,2,...,N(N+1).$ 

The planar limit is obtained by minimizing $V^{trunc}_{eff}$ with respect to the master variables. More precisely, at the minimum, 
\begin{align}
\frac{\partial V^{trunc}_{eff}}{\partial \phi_{\alpha}} \equiv \sum_{C=1}^{N_{\rm lops}}\frac{\partial V^{trunc}_{eff}}{\partial \phi(C)}  \frac{\partial \phi(C) }{\partial \phi_{\alpha}} \Big|_{\phi_{\alpha}^0} &= 0 , \,\, \alpha = 1,2,..., N(N+1) \label{Mmin}\\
\phi_{\rm planar}(C)&\equiv \phi(C)|_{\phi_{\alpha}^0}  , \,\,  C=1,..., N_{\rm loops}.\end{align}
In general, $\partial V^{trunc}_{eff}   / \partial \phi(C) \ne 0$. The planar background is specified by the large $N$ expectation values $\phi_{\rm planar}(C)$ of all gauge invariant operators. 
 
Details of the numerical algorithm have been given in \cite{Koch:2021yeb}. In this article, we have chosen a truncation with $l_{\rm max} = 14$, that is, $2615$ $N_{\rm loops}$ and a $93 \times 93$ $\Omega$ matrix. For the master field, we took $N=51$, corresponding to $2652$ master variables. Evidence for the consistency of the truncation and stability with respect to different values of $N$ has been provided in \cite{Koch:2021yeb}, and is also presented in Appendix A for the Yang-Mills coupled systems considered in this communication.   

\subsection{Spectrum and master variables}

It is important to keep in mind that the $1/N$ expansion is an expansion in terms of loop variables. As such, letting 
\begin{equation*}
\phi(C)=\phi_{\rm planar}(C) + \frac{1}{N} \eta(C) , \hspace{5pt}  \pi(C)=N p(C)   ,    \end{equation*}
we expand (\ref{Hcoll}) up to second order:
\begin{align*}
H^{(2)}_{\rm trunc} &= N \sum_{C=1}^{N_{\rm loops}} \frac{\partial V^{trunc}_{eff}}{\partial \phi(C)}  \eta(C) + \frac{1}{2} \sum_{C,C'=1}^{N_{\rm loops}} p^{\dagger}(C) \hat{\Omega}_0(C,C') p(C') \\
&+ \frac{1}{2} \sum_{C,C'=1}^{N_{\rm loops}} \eta(C) V_0^{(2)} (C,C') \eta^{\dagger}(C') , \hspace{7pt}
V_0^{(2)} (C,C') \equiv \frac{\partial^2 V^{trunc}_{eff}}{\partial \phi(C) \phi^{\dagger}(C')} \Big|_{\phi_{\alpha}^0}\end{align*}
But 
\begin{equation*}\eta(C)= N \sum_{\alpha=1}^{N(N+1)} \frac{\partial \phi(C) }{\partial \phi_{\alpha}} \Big|_{\phi_{\alpha}^0} \, \delta \phi_{\alpha}\end{equation*}
and hence the term linear in $\eta$ in $H^{(2)}_{\rm trunc}$ vanishes, as a result of (\ref{Mmin}).

It is important to note that $\hat{\Omega}_0$ is \underline{not} the same as $\Omega_0$. $\Omega_0$ is a $N_\Omega \times N_\Omega$ matrix ($\Omega$ evaluated at the minimum of $V^{trunc}_{eff}$), but  $\hat{\Omega}_0$ is a $N_{\rm loops} \times N_{\rm loops}$ matrix! In practice, it cannot be calculated in loop space as a loop joining matrix; it would require generating all loops with length $2 l_{\rm max} -2$ or less\footnote{For the truncated system with $l_{\rm max} = 14$, one would have to identify loops containing up to $26$ matrices.}. However, at the minimum, it can be obtained from the planar master field $\phi^0_{\alpha}$ as:
\begin{equation*}
 \hat{\Omega}_0(C,C') =\sum_{A=1}^2 \sum_{a,b=1}^{N}\left( \frac{\partial \phi^{\dagger}(C) }{\partial (X^{\dagger}_A)_{ab}} \right)\Bigg|_{\phi^0_{\alpha}} \left( \frac{\partial \phi(C') }{\partial {(X_A)_ {ba}}} \right) \Bigg|_{\phi^0_{\alpha}}, \, C,C'=1,...,N_{\rm loops}\end{equation*}
A simple analysis of small fluctuations yields for the spectrum eigenvalues:
\begin{equation}\label{eigSp}
\epsilon_n = \left[ \text{eig}_n\left(  \sum_{C'=1}^{N_{\rm loops}}  \hat{\Omega}_0(C,C') V_0^{(2)} (C',C'') \right) \right]^{1/2}\end{equation}

As a result of the difference in dimensions between $ \hat{\Omega}_0$ and $\Omega_0$, there are $N_{\Omega}$ physical, and in general finite, eigenvalues, with $N_{\rm loops}-N_{\Omega}$ zero eigenvalues \cite{Koch:2021yeb, Jevicki:1983hb}
\footnote{It is straightforward to show that the finite eigenvalues of the $2N^2 \times 2N^2$ matrix $  \mathcal{M}_{ab,a'b'} = \sum_{C,C'} \left( \frac{\partial \phi(C') }{\partial {(X_A)_ {ba}}} \right) \Big|_{\phi^0_{\alpha}} V_0^{(2)} (C,C') \left( \frac{\partial \phi^{\dagger}(C) }{\partial (X^{\dagger}_A)_{a'b'}} \right)\Big|_{\phi^0_{\alpha}} $ map to (the square of the) finite eigenvalues of (\ref{eigSp}). We check at every run that the two sets of eigenvalues are identical. Motivation as to why this mass matrix can also be considered is given in\cite{Koch:2021yeb, Jevicki:1983hb}.}.

\section{Massless quantum mechanical system}
In this section we consider the two hermitian matrices ($X_1$ and $X_2$) system Hamiltonian
\begin{equation}\label{FreeHam}
\hat{H} = \frac{1}{2}  \sum_{A=1}^{2} \tr{P_A^2} - \frac{g_{YM}^2}{N} \tr [X_1,X_2]^2 \, . \end{equation}
$P_A$ is canonical conjugate to $X_A\, , A=1,2$.
This system has one dimensional parameter only, $g_{YM}$\footnote{Recall that in terms of our conventions, 't Hooft's coupling $\lambda=g_{YM}^2$.}. Its dimension and that of the fields $X_A$ are:
\begin{equation*}
[g_{YM}]= \frac{3}{2}\, , \hspace{12pt} [X_1]=[X_2]=-\frac{1}{2}\, .
\end{equation*}
 As such, we expect a simple algebraic dependence on $g_{YM}$ of all physical quantities, simply determined by their dimensions. For instance,
 \begin{equation*}
e = \Lambda_e \, g_{YM}^{2/3} \, , \hspace{7pt}  \tr X_1^2  = \Lambda_{[1 1]}  \, g_{YM}^{-2/3} \, ,  \hspace{7pt}  \tr X_1^4  = \Lambda_{[1 1 1 1]}  \, g_{YM}^{-4/3} \, , \hspace{5pt} \rm{etc.},\end{equation*}
where $e$ is any energy of the system.

We considered $15$ values of $g_{YM}$, ranging from $1$ to $12$, chosen so that they are reasonably distributed over this range in both a linear and logarithmic scale, as shown in Table \ref{table:2}:

\begin{table}[h!]
\begin{center}
\resizebox{\textwidth}{!}
{%
\begin{tabular}{||c|c|c|c|c|c|c|c|c|c|c|c|c|c|c||} 
\hline
 \multicolumn{15}{||c||}{$g_{YM}$} \\
\hline\hline
1& 1.28403 & 1.64872& 2& 2.6 & 3.25 & 4 & 5 & 6 & 7 & 8 & 9 & 10 & 11& 12\\ 
 \hline
\end{tabular}
}
\caption{Values of $g_{YM}$. Note that $1.28403\equiv \exp 0.25 \, , 1.64872\equiv \exp 0.50$. }
\label{table:2}
\end{center}
\end{table}
For each value of $g_{YM}$, and directly in this massless limit, we found that the optimization algorithm exhibited remarkable stable convergence to the system's minimum. 
When physical properties are plotted as functions of $g_{YM}$, they show remarkable agreement with their predicted scaling dependence. We present these results in the next subsections, first for large $N$ planar quantities, and then for the spectrum of the theory.

\subsection{Planar limit}

Table \ref{table:3} displays a subset of the results obtained for the planar limit of the quantum mechanical system: the ground state large $N$ energy, the expectation values of all loops with $4$ matrices or less, and an ''angle'', details of which will be given in the following. Following the discussion in Appendix A, we list the ground state energies with $5$ decimal places. Loop data is shown with $4$ decimal places, as to this accuracy loops odd under $1 \to -1$ and $2 \to -2$  vanish, and the symmetry $1 \leftrightarrow 2$ is realized \footnote{Only in very few cases is this not the case, and even in those cases the discrepancy is $\sim 1\times 10^{-4}$.}.

\subsubsection{$U(1)$ charges}
As is known, the system (\ref{FreeHam}) has a global $SO(2) \simeq U(1)$ symmetry associated with rotations of the two hermitian matrices $X_1$ and $X_2$,  with generator 
\begin{equation*}
\hat{L} = -i \, \tr \left( X_1 \frac{\partial}{\partial X_2} -   X_2 \frac{\partial}{\partial X_1} \right). \end{equation*}
In terms of complex matrices $Z\equiv X_1 + i X_2$, $Z^{\dagger }\equiv X_1 - i X_2$, 
\begin{equation*}
Z \to e^{i \phi} Z \, , 
\hspace{6pt}
Z^{\dagger} \to e^{- i \phi} Z^{\dagger} \, , \hspace{6pt}\hat{L} = \ \tr \left( Z \frac{\partial}{\partial Z} -   Z^\dagger \frac{\partial}{\partial Z^\dagger} \right)
\, .
\end{equation*}
At the planar level, expectation values of correlators with non-zero $U(1)$ charges should vanish. In order to evidence this, we display the expectation values of all loops with $4$ complex matrices or less in table \ref{table:3B}. 
The vanishing of all listed non-zero charge planar correlators is realized to within at least $5$ significant figures \footnote{We do not display the imaginary part of the generically complex loops, as these are numerically zero to at least $4$ decimal places for the loops presented in table \ref{table:3B}. The fact that their imaginary part is numerically zero follows from the vanishing of loops in table \ref{table:3} that are odd under $1 \to -1$ or $2 \to -2$. }.  

\subsubsection{Scaling behaviour}

\begin{figure}[h!]
\centering
\includegraphics[width=8cm]{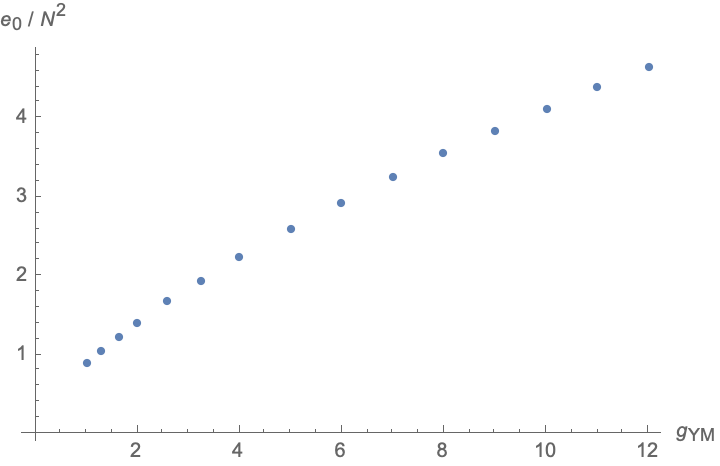}
\caption{Planar ground state energies $e_0/N^2$ versus $g_{YM}$ }
\label{fig:1-E0-vs-g_YM}
\end{figure}

\begin{figure}[h!]
\centering
\includegraphics[width=8cm]{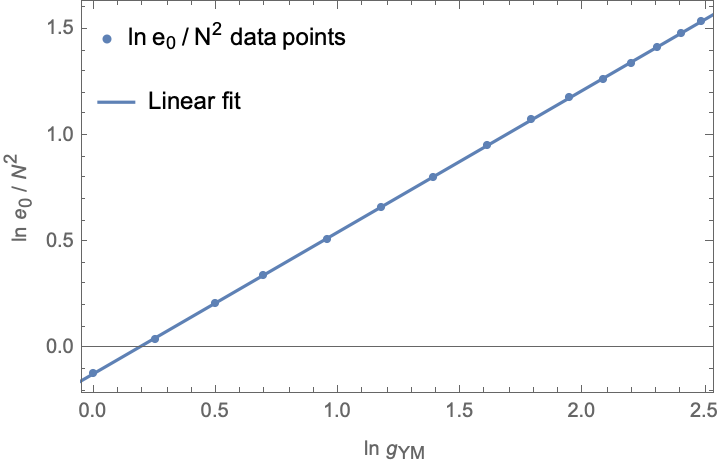}
\caption{Linear fit to $\ln e_0/N^2$ versus $\ln g_{YM}$ }
\label{fig:2-Linear E0-vs-g_YM}
\end{figure}

In figure \ref{fig:1-E0-vs-g_YM}, a plot of large $N$ ground state energies versus $g_{YM}$ is shown.
We fit the data to the curve
\begin{equation*}
e_0 / N^2 = {A}_0 \,  g_{YM}^p \, ,
\end{equation*}
by performing a linear regression (least squares) fit to the logarithmic plot. We find:
\begin{equation*}
\ln {A}_0 = -0.117625(8) \, , \hspace{10pt} p = 0.666671(5) \, .  \end{equation*}
This linear fit is shown in Figure \ref{fig:2-Linear E0-vs-g_YM}.

\begin{sidewaystable}
\begin{center}
\resizebox{1.1\textwidth}{!}{%
\begin{tabular}{||c|c|c|c|c|c|c|c|c|c|c|c|c|c|c|c||} 
\hline 
$g_{YM}$ & 1.0 &  1.28403& 1.64872& 2.0 & 2.6 & 3.25 & 4.0 & 5.0 & 6.0 & 7.0 & 8.0 & 9.0 & 10.0 & 11.0 & 12.0\\
\hline 
$e_{0}/N^{2}$ & 0.88904 & 1.05027 & 1.24075 & 1.41124 & 1.68099 & 1.95060 & 2.24021 & 2.59951 & 2.93549 & 3.25320 & 3.55622 & 3.84669 & 4.12662 & 4.39716 & 4.65995\\
\hline 
$\tr \, {1} /N$ &  1.0000 & 1.0000 & 1.0000 & 1.0000 & 1.0000 & 1.0000 & 1.0000 & 1.0000 & 1.0000 & 1.0000 & 1.0000 & 1.0000 & 1.0000 & 1.0000 & 1.0000\tabularnewline
\hline 
$[1]\, /N^{3/2}$ &  0.0000 & 0.0000 & 0.0000 & 0.0000 & 0.0001 & 0.0000 & 0.0000 & 0.0000 & 0.0000 & 0.0000 & -0.0001 & 0.0000 & -0.0001 & 0.0000 & 0.0000\tabularnewline
\hline 
$[2]\,/N^{3/2}$ &  0.0000 & 0.0000 & 0.0003 & 0.0000 & 0.0000 & 0.0000 & 0.0000 & 0.0000 & 0.0000 & 0.0000 & 0.0000 & 0.0000 & 0.0000 & 0.0000 & 0.0000\tabularnewline
\hline 
$[11]\, /N^{2}$ &  0.4651 & 0.3937 & 0.3333 & 0.2931 & 0.2460 & 0.2120 & 0.1846 & 0.1591 & 0.1409 & 0.1271 & 0.1163 & 0.1075 & 0.1002 & 0.0941 & 0.0887\tabularnewline
\hline 
$[12]\, /N^{2}$ &  0.0000 & 0.0000 & 0.0000 & 0.0000 & 0.0000 & 0.0000 & 0.0000 & 0.0000 & 0.0000 & 0.0000 & 0.0000 & 0.0000 & 0.0000 & 0.0000 & 0.0000\tabularnewline
\hline 
$[22]\, /N^{2}$ &  0.4651 & 0.3937 & 0.3333 & 0.2930 & 0.2460 & 0.2120 & 0.1846 & 0.1591 & 0.1409 & 0.1271 & 0.1163 & 0.1075 & 0.1002 & 0.0941 & 0.0887\tabularnewline
\hline 
$[111]\, /N^{5/2}$ &  0.0000 & 0.0000 & 0.0001 & 0.0000 & 0.0001 & 0.0000 & 0.0000 & 0.0000 & 0.0000 & 0.0000 & -0.0001 & 0.0000 & 0.0000 & 0.0000 & 0.0000\tabularnewline
\hline 
$[112]\, /N^{5/2}$ &  0.0000 & 0.0000 & 0.0001 & 0.0000 & 0.0000 & 0.0000 & 0.0000 & 0.0000 & 0.0000 & 0.0000 & 0.0000 & 0.0000 & 0.0000 & 0.0000 & 0.0000\tabularnewline
\hline 
$[122]\, /N^{5/2}$ &  0.0000 & 0.0000 & 0.0000 & 0.0000 & 0.0000 & 0.0000 & 0.0000 & 0.0000 & 0.0000 & 0.0000 & 0.0000 & 0.0000 & 0.0000 & 0.0000 & 0.0000\tabularnewline
\hline 
$[222]\, /N^{5/2}$ &  -0.0001 & 0.0000 & 0.0003 & 0.0000 & 0.0000 & 0.0000 & 0.0000 & 0.0000 & 0.0000 & 0.0000 & 0.0000 & 0.0000 & 0.0000 & 0.0000 & 0.0000\tabularnewline
\hline 
$[1111]\, /N^{3}$ &  0.4364 & 0.3127 & 0.2241 & 0.1732 & 0.1221 & 0.0907 & 0.0687 & 0.0511 & 0.0400 & 0.0326 & 0.0273 & 0.0233 & 0.0203 & 0.0178 & 0.0159\tabularnewline
\hline 
$[1112]\, /N^{3}$ &  0.0000 & 0.0000 & 0.0000 & 0.0000 & 0.0000 & 0.0000 & 0.0000 & 0.0000 & 0.0000 & 0.0000 & 0.0000 & 0.0000 & 0.0000 & 0.0000 & 0.0000\tabularnewline
\hline 
$[1122]\, /N^{3}$ &  0.1949 & 0.1396 & 0.1001 & 0.0773 & 0.0545 & 0.0405 & 0.0307 & 0.0228 & 0.0179 & 0.0146 & 0.0122 & 0.0104 & 0.0090 & 0.0080 & 0.0071\tabularnewline
\hline 
$[1212]\, /N^{3}$ &  0.0467 & 0.0334 & 0.0240 & 0.0185 & 0.0131 & 0.0097 & 0.0074 & 0.0055 & 0.0043 & 0.0035 & 0.0029 & 0.0025 & 0.0022 & 0.0019 & 0.0017\tabularnewline
\hline 
$[1222]\, /N^{3}$ &  0.0000 & 0.0000 & 0.0000 & 0.0000 & 0.0000 & 0.0000 & 0.0000 & 0.0000 & 0.0000 & 0.0000 & 0.0000 & 0.0000 & 0.0000 & 0.0000 & 0.0000\tabularnewline
\hline 
$[2222]\, /N^{3}$ &  0.4364 & 0.3127 & 0.2241 & 0.1732 & 0.1221 & 0.0907 & 0.0687 & 0.0511 & 0.0400 & 0.0326 & 0.0273 & 0.0233 & 0.0202 & 0.0178 & 0.0159\tabularnewline
\hline 
"angle" & 0.68489 & 0.68488 & 0.68482 & 0.68478 & 0.68477 & 0.68472 & 0.68474 & 0.68472 & 0.68472 & 0.68471 & 0.68518 & 0.68513 & 0.68516 & 0.68466 & 0.68516\tabularnewline
\hline 
\end{tabular}}
\caption{Planar numerical results obtained with a truncation to $2615$ loops ($l_{\rm max} =14$) with $\Omega$ a $93 \times 93$ matrix. }
\label{table:3}
\end{center}
\end{sidewaystable}

\begin{sidewaystable}
\begin{center}
\resizebox{1.1\textwidth}{!}{%
\begin{tabular}{||c|c|c|c|c|c|c|c|c|c|c|c|c|c|c|c|c||} 
\hline 
 &$g_{YM}$ & 1,00 & 1,28 & 1,65 & 2,00 & 2,60 & 3,25 & 4,00 & 5,00 & 6,00 & 7,00 & 8,00 & 9,00 & 10,00 & 11,00 & 12,00  \\
\hline
$l$& Energy & 0.88904 & 1.05027 & 1.24075 & 1.41124 & 1.68099 & 1.95060 & 2.24021 & 2.59951 & 2.93549 & 3.25320 & 3.55622 & 3.84669 & 4.12662 & 4.39716 & 4.65995  \\
\hline
0& $\mathrm{Tr} 1/N$ &  1.0000 & 1.0000 & 1.0000 & 1.0000 & 1.0000 & 1.0000 & 1.0000 & 1.0000 & 1.0000 & 1.0000 & 1.0000 & 1.0000 & 1.0000 & 1.0000 & 1.0000  \\
\hline
1&$[Z]/N^{3/2}$ & 0.0000 & 0.0000 & 0.0000 & 0.0000 & 0.0000 & -0.0000 & 0.0000 & 0.0000 & 0.0000 & -0.0001 & -0.0000 & -0.0000 & -0.0000 & -0.0000 & 0.0000  \\
\hline
-1&$[Z^\dagger]/N^{3/2}$ &0.0000 & 0.0000 & 0.0000 & 0.0000 & 0.0000 & -0.0000 & 0.0000 & 0.0000 & 0.0000 & -0.0001 & -0.0000 & -0.0000 & -0.0000 & -0.0000 & 0.0000 \\
\hline
2&$[ZZ]/N^{2}$ & 0.0000 & 0.0000 & -0.0000 & 0.0000 & -0.0000 & -0.0000 & 0.0000 & -0.0000 & -0.0000 & 0.0000 & -0.0000 & 0.0000 & 0.0000 & -0.0000 & 0.0000 \\
\hline
0&$[ZZ^\dagger]/N^{2}$ & 0.9303 & 0.7874 & 0.6665 & 0.5861 & 0.4920 & 0.4240 & 0.3692 & 0.3182 & 0.2818 & 0.2542 & 0.2325 & 0.21500 & 0.2004 & 0.1881 & 0.1774  \\
\hline
-2&$[Z^\dagger Z^\dagger]/N^{2}$ & 0.0000 & 0.0000 & -0.0000 & 0.0000 & -0.0000 & -0.0000 & 0.0000 & -0.0000 & -0.0000 & 0.0000 & 0.0000 & 0.0000 & -0.0000 & 0.0000 & 0.0000  \\
\hline
3&$[Z^3]/N^{5/2}$ & 0.0000 & 0.0000 & 0.0000 & 0.0000 & -0.0000 & 0.0000 & -0.0000 & -0.0000 & 0.0000 & -0.0000 & -0.0000 & -0.0000 & 0.0000 & 0.0000 & -0.0000  \\
\hline
1&$[ZZZ^\dagger]//N^{5/2}$& 0.0000 & 0.0000 & 0.0000 & 0.0000 & 0.0000 & -0.0000 & 0.0000 & 0.0000 & 0.0000 & 0.0000 & -0.0000 & -0.0000 & -0.0000 & -0.0000 & -0.0000  \\
\hline
-1&$[ZZ^\dagger Z^\dagger]/N^{5/2}$&  0.0000 & 0.0000 & 0.0000 & 0.0000 & 0.0000 & -0.0000 & 0.0000 & 0.0000 & 0.0000 & 0.0000 & -0.0000 & -0.0000 & -0.0000 & -0.0000 & -0.0000  \\
\hline
-3&$[Z^\dagger Z^\dagger Z^\dagger]/N^{5/2}$&  0.0000 & 0.0000 & 0.0000 & 0.0000 & -0.0000 & 0.0000 & -0.0000 & -0.0000 & 0.0000 & -0.0000 & -0.0000 & -0.0000 & 0.0000 & 0.0000 & -0.0000  \\
\hline
4&$[ZZZZ]/N^{3}$ & 0.0000 & 0.0000 & 0.0000 & -0.0000 & 0.0000 & -0.0000 & 0.0000 & -0.0000 & -0.0000 & -0.0000 & -0.0000 & 0.0000 & 0.0000 & 0.0000 & -0.0000  \\
\hline
2&$[ZZZZ^\dagger]/N^{3}$ & 0.0000 & 0.0000 & -0.0000 & 0.0000 & 0.0000 & 0.0000 & 0.0000 & 0.0000 & -0.0000 & -0.0000 & -0.0000 & -0.0000 & 0.0000 & 0.0000 & 0.0000 \\
\hline
0&$[ZZZ^\dagger Z^\dagger]/N^{3}$&  0.9662 & 0.6923 & 0.4961 & 0.3835 & 0.2703 & 0.2008 & 0.1522 & 0.1130 & 0.0886 & 0.0722 & 0.0604 & 0.0515 & 0.0448 & 0.0395 & 0.0352 \\
\hline
0&$[ZZ^\dagger ZZ^\dagger]/N^{3}$ & 1.5589 &	1.11698 &	0.8004 &	0.6187 &	0.4361 &	0.3239 &	0.2456 &	0.1824 &	0.1430 &	0.1164 &	0.0974 &	0.0832 &	0.0723 &	0.0637 &	0.0567 \\
\hline
-2&$[Z Z^\dagger Z^\dagger Z^\dagger]/N^{3}$ & 0.0000 & 0.0000 & -0.0000 & 0.0000 & 0.0000 & 0.0000 & 0.0000 & 0.0000 & -0.0000 & -0.0000 & -0.0000 & -0.0000 & 0.0000 & 0.0000 & 0.0000 \\
\hline
-4&$[Z^\dagger Z^\dagger Z^\dagger Z^\dagger]/N^{3}$ & 0.0000 & 0.0000 & -0.0000 & 0.0000 & 0.0000 & 0.0000 & 0.0000 & 0.0000 & -0.0000 & -0.0000 & -0.0000 & -0.0000 & 0.0000 & 0.0000 & 0.0000 \\
\hline
\end{tabular}}
\caption{Real part of planar expectation values of correlators built with complex matrices and $U(1)$ charges $l$. }
\label{table:3B}
\end{center}
\end{sidewaystable}

The accuracy with which the interpolation matches the exact scaling $p=2/3$ at this level of truncation is remarkable \footnote{The parameters and their uncertainties are obtained with the Mathematica functions LinearModelFit and NonlinearModelFit. Note that the fit uncertainties are remarkably consistent with the estimated numerical errors associated with this level of loop truncation.}. We are then justified in setting $p=2/3$ and fit the data to the scaling function
\begin{equation}\label{E0Scaling}
e_0 / N^2 = \Lambda_0 \,  g_{YM}^{2/3}\, , \hspace{5pt} \left(= \Lambda_0 \,  \lambda^{1/3} \right) \end{equation}
with result 
\begin{equation}\label{E0ScPar}
\Lambda_0=0.889034(3) .
\end{equation}

\begin{figure}[h!]
\centering
\includegraphics[width=8cm]{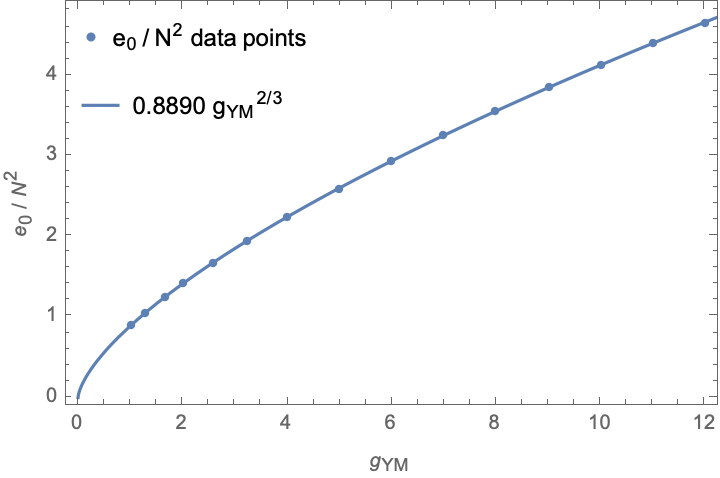}
\caption{Numerically obtained large $N$ ground state energies and fit to the scaling function $e_0 / N^2 = 0.8890 \,  g_{YM}^{2/3}$ }
\label{fig:3-Final E0-vs-g_YM}
\end{figure}

Figure \ref{fig:3-Final E0-vs-g_YM} displays the fit of the large $N$ planar ground state energies to the scaling function (\ref{E0Scaling}) with parameter (\ref{E0ScPar}). Again, the level of accuracy (\ref{E0ScPar}) with which the numerically obtained planar ground state energies match the scaling behaviour (\ref{E0Scaling}) at this level of truncation is remarkable. 

Taking into account possible truncation dependent errors, as estimated in Appendix A, we then list the final scaling dependence on 't Hooft's coupling for the planar ground state energy of the massless system as:

\begin{equation*}
\boxed{
e_0 / N^2=0.8890(2) \,  \lambda^{1/3}  
}
\end{equation*}

\begin{figure}[h!]
    \centering
    \subfloat[Linear fit of $\ln \tr Z^{\dagger}Z/N^2$ versus $\ln g_{YM}$]{{\includegraphics[width=0.45\textwidth]{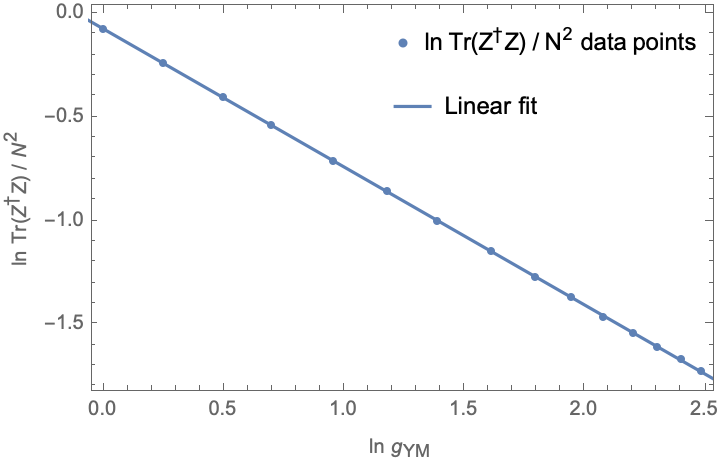} }}
    \qquad
    \subfloat[Fit of $\tr Z^{\dagger}Z/N^2$ to scaling function $ 0.9303 \, \, g_{YM}^{-2/3} $ ]{{\includegraphics[width=0.45\textwidth]{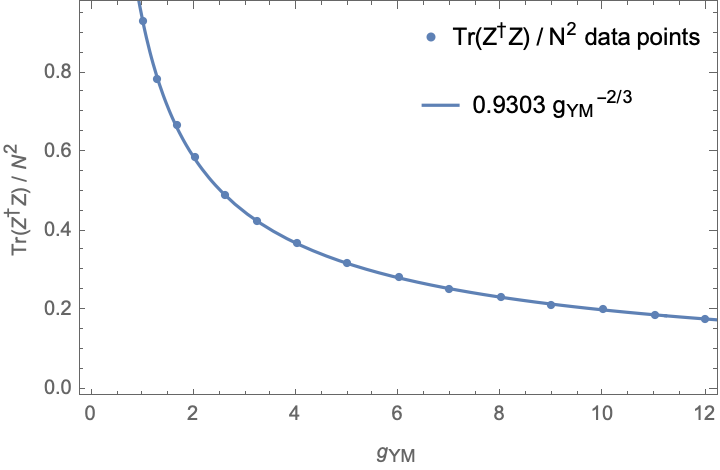} }}
    \caption{Numerical results for the planar limit of $\tr Z^{\dagger}Z/N^2$, logarithmic linear fit and fit to predicted scaling dependence. }
    \label{fig:4-ZZ-vs-g_YM}
\end{figure}

We follow the same analysis for loops with two matrices and consider the correlator $\tr (Z^{\dagger} Z) /N^2 = (\tr X_1^2 + \tr X_2^2)/N^2$. The logarithmic $g_{YM}$ dependence is first approximated by a linear fit, and then matched to the scaling dimensions of the loop correlator:
\begin{equation*}
\tr (Z^{\dagger} Z) /N^2 = {A}_{Z^{\dagger} Z} \,  g_{YM}^p \, , \hspace{8pt} \text{and then} \hspace{5pt} \tr (Z^{\dagger} Z) /N^2 = \Lambda_{Z^{\dagger} Z} \, \, g_{YM}^{-2/3} \, .\end{equation*}
The results are presented in table \ref{table:4} and displayed in Figure \ref{fig:4-ZZ-vs-g_YM}.
\begin{table}[h!]
\begin{center}
\begin{tabular}{|| c | c || c ||c||} 
\hline
\multicolumn{2}{||c||}{Parameters of (log) linear fit} &{ $p = -2/3$} fixed & {Final scaling function} \\\hline\hline
$\ln {A}_{Z^{\dagger} Z}$& $p$ & $\Lambda_{Z^{\dagger} Z}$ & $\tr (Z^{\dagger} Z) /N^2 $  \\ 
\hline
-0.07219(7) & -0.66672(4) &      0.93027(3) &     { $ 0.930(1) \, \lambda^{-1/3}$}\\ 
 \hline
\end{tabular}
\caption{$\tr (Z^{\dagger}Z)/N^2$  log linear fit parameters and scaling parameter $\Lambda_{Z^{\dagger} Z}$ at this level of truncation. The scaling parameter in the final scaling function takes into account estimated effects of the loop truncation.  }
\label{table:4}
\end{center}
\end{table}

The scaling power for the large $N$ planar correlator is again predicted with a high level of accuracy, and their numerical values match with a high level of precision the scaling behaviour. The numerical errors associated with the loop truncation are estimated in Appendix A, and taken into account in the final scaling function presented in table \ref{table:4}.

\begin{figure}[h!]
    \centering
    \subfloat[Linear fit of the log of $4$ matrices loop expectation values versus $\ln g_{YM}$]{{\includegraphics[width=0.45\textwidth]{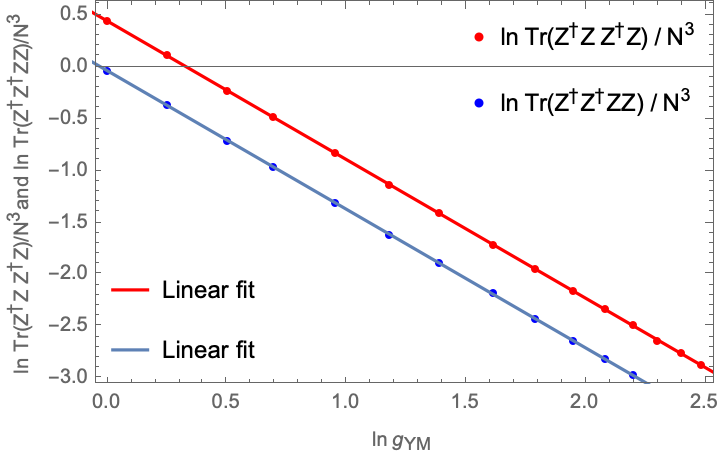} }}
    \qquad
    \subfloat[Fit of loops of $4$ matrices to scaling functions $ \Lambda \, \, g_{YM}^{-4/3} $ ]{{\includegraphics[width=0.45\textwidth]{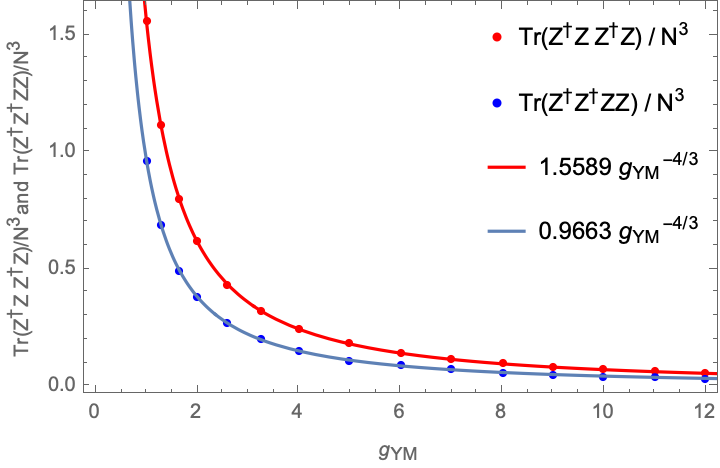} }}
    \caption{Numerical results for the planar limit of $\tr (Z^{\dagger} Z Z^{\dagger} Z) /N^3$ and $\tr (Z^{\dagger} Z^{\dagger} Z Z)/N^3$, logarithmic linear fits and fits to predicted scaling dependence.}
    \label{fig:5-ZZZZ-vs-g_YM}
\end{figure}

For invariant loops with $4$ matrices, we consider the loops $\tr (Z^{\dagger} Z Z^{\dagger} Z) /N^3$ and  $\tr (Z^{\dagger} Z^{\dagger} Z Z)/N^3$, and carry out the same analysis, which is summarized in table \ref{table:5} and figure \ref{fig:5-ZZZZ-vs-g_YM}.
\begin{table}[h!]
\begin{center}
\begin{tabular}{||c||c|c||c||c||} 
\hline
& \multicolumn{2}{|c||}{Log linear fit} & {$p=-4/3$} & Final\\
\hline\hline
 &$\ln {A}$& $p$ & $\Lambda$ & Scaling function \\ 
\hline
$\tr (Z^{\dagger} Z Z^{\dagger} Z) /N^3$  &0.4441(1) & -1.33340(6)  &    1.55895(8)  & $1.559(8) \, \lambda^{-2/3}$ \\ 
 \hline
 $\tr (Z^{\dagger} Z^{\dagger} Z Z)/N^3$ &-0.0342(2)  & -1.3334(1)  &    0.96626(8) & $0.966(4)  \, \lambda^{-2/3}$ \\ 
\hline
\end{tabular}
\caption{Logarithmic linear fit parameters and scaling parameter for $\tr (Z^{\dagger} Z Z^{\dagger} Z) /N^3$ and $\tr (Z^{\dagger} Z^{\dagger} Z Z)/N^3$. }
\label{table:5}
\end{center}
\end{table}

Remarks similar to those given for the previously discussed large $N$ planar quantities which concern the high level of accuracy of the numerical results, apply to the invariant loops with $4$ matrices considered in this case too.    

Finally, we consider an "angle" defined to be 
\begin{equation*}
\mathcal{A} \equiv  N \frac{\tr X_1^2 X_2^2 - \tr X_1X_2X_1X_2}{\tr {X_1^2} \tr X_2^2} = -\frac{N}{2} \frac{ \tr [X_1,X_2]^2 }{\tr {X_1^2} \tr X_2^2} \, .
\end{equation*}
For the integral with masses, and at large coupling, it has been shown that the two matrices commute \cite{Berenstein:2008eg}. For the quantum mechanical system in the massless limit, we observe, from the data listed in \ref{table:3} that this ratio remains constant, and we obtain
\begin{equation*}
\mathcal{A} = 0.68487(5) \, .
\end{equation*}
Again, taking into account possible numerical errors associated with the loop truncation, we obtain 
\begin{equation*}\boxed{
\mathcal{A} = N \frac{\tr X_1^2 X_2^2 - \tr X_1X_2X_1X_2}{\tr {X_1^2} \tr X_2^2}  = 0.685(2) \, .}\end{equation*}

In \cite{Han:2020bkb}, the large $N$ YM quantum mechanics of two matrices with mass was considered, and it was there pointed out that this ratio seemed to show convergence to a constant value at large coupling. We have been able to establish what its value is directly in the massless limit, for all coupling values, with a high degree of accuracy.

\subsection{$1/N$ spectrum }
We consider in this subsection the numerical results obtained for the spectrum of the theory. These are independent of $N$ and are determined from the quadratic hamiltonian $H^{(2)}_{\rm trunc}$ as $1/N$ fluctuations about the large $N$ planar background, as described in Section 2.3.

\subsubsection{Masses and scaling behaviour}
We observe that the mass of the third excited state and of all other higher excited states show the expected increase with coupling. The same is not the case for the two lowest lying states. We will first concentrate on the third and higher excited states, and discuss the two lowest lying states at the end of this subsection.

Numerically, one finds that the lowest lying states are determined quite accurately for small loop truncations, with the next higher lying states then becoming more accurate with larger number loops, and so on.

Table \ref{table:6} displays the numerical results obtained for the masses of the $3$rd to the $37$th excited state \footnote{These would be $2$ particle states to $6$ particle states in the free system with a mass coupling.} as a function of the coupling constant $g_{YM}$. Numerically, one finds that the lowest lying states are determined quite accurately for small loop truncations, with the next higher lying states then becoming more accurate with a larger number loops, and so on. Although the code generates data for all first $93$ excited states, we have chosen to list masses up to the $37$th excited state as their masses are stable at this level of truncation.  Truncation error estimates are presented in Appendix A.  

We then carry out a similar analysis to that of previous subsection, by first performing a linear fit to the dependence of $\ln e_n , n=3,...,37$ on the logarithm of $g_{YM}$, comparing it with the the scaling power prediction, and then optimize the match to the scaling dependence of the energies:
\begin{equation*}
e_n = {A}_{n} \,  g_{YM}^p \, , \hspace{8pt} \text{and then} \hspace{5pt} e_n = \Lambda_{n} \, \, g_{YM}^{2/3} \, .\end{equation*}
The  $\Lambda_{n}$ are then adjusted to take into account truncation error estimates. The results are presented in table \ref{table:7}\footnote{The somewhat unconventional notation $4.62(21)$ say, instead of $4.6(2)$, is meant to help compare final parameter error estimates with those of the current ($2615$ loops) truncation level (in this case $4.62(2)$) and highlight possible differences within possible multiplets.}. 

\begin{sidewaystable}
\begin{center}
\resizebox{1.1\textwidth}{!}{%
\begin{tabular}{||c|c|c|c|c|c|c|c|c|c|c|c|c|c|c|c||} 
\hline 
$g_{YM}$ & 1.0 &  1.28403& 1.64872& 2.0 & 2.6 & 3.25 & 4.0 & 5.0 & 6.0 & 7.0 & 8.0 & 9.0 & 10.0 & 11.0 & 12.0\\
\hline 
\hline
$e_3$ &1.58783&	1.87590&	2.21627&	2.52042&	3.00208&	3.48335&	4.00097&	4.64009&	5.24314&	5.80891&	6.35076&	6.87178&	7.36991&	7.84974	&8.32061\tabularnewline
\hline
$e_4$& 1.58826&	1.87666&	2.21658&	2.52084&	3.00301&	3.48399&	4.00108&	4.64333&	5.24456&	5.80914&	6.35230&	6.87266&	7.37212&	7.85155&	8.32270\tabularnewline
\hline
$e_5$& 1.8817	&      2.2384&	2.6469&	2.9897&	3.5754&	4.1113&	4.7403&	5.4561&	6.1860&	6.8548&	7.2816&	7.8399&	8.4051&	9.1596&	9.5306\tabularnewline
\hline
$e_6$& 2.3624	&      2.8549&	3.3788&	3.6678&	4.5283&	5.2501&	6.0686&	6.9439&	7.8667&	8.6501&	9.4171&	10.217&	10.864&	11.630&	12.340\tabularnewline
\hline
$e_7$& 2.4227	&      2.8614&	3.3845&	3.8005&	4.5699&	5.2754&	6.0982&	7.0592&	8.0208&	8.7401&	9.5187&	10.368&	11.175&	11.731&	12.574\tabularnewline
\hline
$e_8$& 2.8920	&      3.5499&	4.2006&	4.5165&	5.6054&	6.4257&	7.4950&	8.5833&	9.7959&	10.620&	11.343&	12.400&	13.159&	14.118&	14.850\tabularnewline
\hline
$e_9$& 3.0224	&      3.5679&	4.2135&	4.7529&	5.6196&	6.5906&	7.6095&	8.7697&	9.9158&	10.999&	11.722&	12.656&	13.328&	14.740&	15.106\tabularnewline
\hline
$e_{10}$& 3.1770&	3.7796&	4.4787&	4.9936&	6.0322&	6.9061&	8.0139&	9.2056&	10.538&	11.536&	12.659&	13.720&	14.655&	15.607&	16.529\tabularnewline
\hline
$e_{11}$& 3.2119&	3.7908&	4.4854&	5.0488&	6.0746&	7.0056&	8.0833&	9.2782&	10.605&	11.620&	12.732&	13.763&	14.762&	15.718&	16.576\tabularnewline
\hline
$e_{12}$& 3.6644&	4.3828&	5.2722&	5.7622&	7.0210&	8.1770&	9.3819&	10.537&	12.173&	13.366&	13.179&	14.243&	15.329&	17.050&	17.522\tabularnewline
\hline
$e_{13}$& 3.7820&	4.5010&	5.3748&	5.9431&	7.2252&	8.2713&	9.4737&	10.789&	12.494&	13.589&	15.134&	16.433&	17.470&	18.407&	19.565\tabularnewline
\hline
$e_{14}$& 3.9256&	4.5641&	5.4285&	6.0557&	7.3141&	8.4786&	9.8013&	11.268&	12.813&	13.813&	15.230&	16.737&	17.774&	18.540&	19.990\tabularnewline
\hline
$e_{15}$& 3.9768&	4.6918&	5.5272&	6.1198&	7.5254&	8.6510&	10.060&	11.526&	13.202&	14.089&	15.555&	16.916&	18.021&	18.930&	20.128\tabularnewline
\hline
$e_{16}$& 3.9904&	4.7478&	5.6144&	6.3157&	7.5563&	8.7746&	10.124&	11.600&	13.250&	14.597&	15.666&	16.972&	18.166&	19.639&	20.415\tabularnewline
\hline
$e_{17}$& 4.2782&	5.1428&	6.0890&	6.7240&	8.1507&	9.1677&	10.767&	12.325&	13.917&	15.308&	17.491&	18.728&	20.168&	20.714&	22.509\tabularnewline
\hline
$e_{18}$& 4.7181&	5.5377&	6.6292&	7.1827&	8.8371&	10.081&	11.790&	13.342&	15.513&	16.652&	18.382&	19.823&	21.356&	22.874&	23.684\tabularnewline
\hline
$e_{19}$& 4.7579&	5.6074&	6.6581&	7.4679&	8.9802&	10.353&	11.879&	13.656&	15.716&	17.144&	18.613&	20.155&	21.445&	23.092&	24.225\tabularnewline
\hline
$e_{20}$& 4.7994&	5.6289&	6.6876&	7.5628&	9.1130&	10.473&	12.018&	13.895&	15.768&	17.276&	18.952&	20.882&	22.208&	23.219&	25.016\tabularnewline
\hline
$e_{21}$& 4.8160&	5.6543&	6.7112&	7.6996&	9.1279&	10.549&	12.048&	13.920&	15.935&	17.511&	19.144&	20.921&	22.451&	23.420&	25.119\tabularnewline
\hline
$e_{22}$& 5.0709&	5.8664&	7.0180&	7.9112&	9.4813&	11.056&	12.703&	14.521&	16.572&	18.178&	19.980&	21.713&	23.141&	24.453&	26.087\tabularnewline
\hline
$e_{23}$& 5.1189&	5.9758&	7.1513&	8.0451&	9.6420&	11.162&	12.899&	14.915&	16.728&	18.454&	20.172&	21.979&	23.337&	24.548&	26.198\tabularnewline
\hline
$e_{24}$& 5.5177&	6.4132&	7.7126&	8.6924&	10.400&	11.973&	13.669&	15.655&	18.175&	19.692&	21.818&	23.843&	25.621&	26.910&	28.568\tabularnewline
\hline
$e_{25}$& 5.5663&	6.5097&	7.7646&	8.8801&	10.577&	12.156&	14.062&	15.994&	18.407&	20.141&	22.037&	24.103&	25.747&	27.387&	29.097\tabularnewline
\hline
$e_{26}$& 5.6155&	6.6611&	7.8974&	8.8917&	10.701&	12.409&	14.195&	16.348&	18.606&	20.372&	22.472&	24.597&	26.309&	27.733&	29.533\tabularnewline
\hline
$e_{27}$& 5.6549&	6.7057&	7.9368&	8.9654&	10.722&	12.487&	14.344&	16.464&	18.706&	20.675&	22.619&	24.643&	26.417&	27.766&	29.735\tabularnewline
\hline
$e_{28}$& 5.9641&	6.8887&	8.2739&	9.4674&	11.256&	12.937&	14.844&	16.996&	19.266&	21.269&	23.800&	25.686&	27.749&	29.054&	31.282\tabularnewline
\hline
$e_{29}$& 6.0821&	7.0254&	8.3083&	9.4841&	11.329&	13.063&	15.009&	17.416&	19.623&	21.800&	24.063&	26.090&	27.884&	29.207&	31.400\tabularnewline
\hline
$e_{30}$& 6.1661&	7.1322&	8.5372&	9.6920&	11.623&	13.387&	15.364&	17.831&	20.078&	22.169&	24.848&	26.217&	28.152&	29.996&	31.865\tabularnewline
\hline
$e_{31}$& 6.3174&	7.5318&	8.8912&	9.9197&	11.876&	13.525&	15.821&	18.185&	20.581&	22.576&	25.696&	28.703&	30.412&	30.253&	34.036\tabularnewline
\hline
$e_{32}$& 6.4214&	7.7110&	9.0560&	10.269&	12.467&	14.282&	16.409&	18.871&	21.506&	23.665&	26.221&	28.788&	30.709&	32.035&	34.632\tabularnewline
\hline
$e_{33}$& 6.4975&	7.7573&	9.0860&	10.373&	12.514&	14.618&	16.617&	19.174&	21.600&	23.724&	26.621&	29.246&	31.143&	32.371&	35.148\tabularnewline
\hline
$e_{34}$& 6.7837&	8.0036&	9.4812&	10.756&	12.904&	14.845&	16.954&	19.663&	22.339&	24.689&	26.737&	29.415&	31.332&	33.115&	35.432\tabularnewline
\hline
$e_{35}$& 6.7997&	8.0186&	9.4927&	10.793&	12.944&	14.877&	17.047&	19.847&	22.377&	24.816&	27.003&	29.730&	31.483&	33.359&	35.767\tabularnewline
\hline
$e_{36}$& 6.9259&	8.0801&	9.6280&	10.935&	13.037&	15.090&	17.302&	19.985&	22.726&	24.963&	27.355&	29.820&	32.145&	33.537&	36.135\tabularnewline
\hline
$e_{37}$&7.0077&	8.2133&	9.7493&	11.030&	13.209&	15.377&	17.472&	20.177&	23.011&	25.253&	27.864&	29.948&	32.250&	34.672&	36.447\\
\hline 
\end{tabular}}
\caption{Numerical results for the spectrum with a truncation to $2615$ loops ($l_{\rm max} =14$) with $\Omega$ a $93 \times 93$ matrix. Only the states $n=3,...,37$ are listed.}
\label{table:6}
\end{center}
\end{sidewaystable}

\begin{table}[h!]
\begin{center}
\begin{tabular}{||c||c|c||c||c||} 
\hline
& \multicolumn{2}{|c||}{Log linear fit} & $p=2/3$ fixed& Final\\
\hline\hline
n &$\ln {A}_n $& $p$ & $\Lambda_n$ & Scaling function  \\ 
\hline
$e_{3}$   &0.4624(1)  & 0.66657(7)  &  1.58767(9)     & $1.588(1) \, \, \lambda^{1/3}$ \\ 
 \hline
$e_{4}$ & 0.4627(1) &0.66656(6)   &  1.58806(8)    & $1.588(1)\,\, \lambda^{1/3}$ \\ 
  \hline
$e_{5}$  &0.645(6)     & 0.650(3)   &  1.862(8)   &$1.86(3)\,\, \lambda^{1/3}$  \\
  \hline
$e_{6}$ &  0.873(6)& 0.660(4)  &   2.373(8)   & $2.37(3)\,\, \lambda^{1/3}$  \\
  \hline
 $e_{7}$ & 0.885(3) &0.661(2) & 2.406(5)     & $2.41(3)\,\, \lambda^{1/3}$  \\
  \hline
 $e_{8}$  &1.09(1)   & 0.651(6)   &  2.91(2)    & $2.91(11)\,\, \lambda^{1/3}$  \\
  \hline
 $e_{9}$ &1.112(7)   & 0.652(4)  &   2.98(1)  & $2.98(10)\, \lambda^{1/3}$ \\
  \hline
 $e_{10}$ &1.159(3)  & 0.663(2)  &   3.170(5)   & $3.17(2)\,\, \lambda^{1/3}$  \\
  \hline
 $e_{11}$&1.167(2)  &  0.662(1)  & 3.191(5)     & $3.19(2)\, \lambda^{1/3}$ \\
  \hline
 $e_{12}$ &1.34(2)  &  0.62(1)  &  3.57(5)    & $3.57(18)\,\, \lambda^{1/3}$  \\
  \hline
$e_{13}$ & 1.336(6)& 0.660(3)   &   3.77(1)  & $3.77(6)\,\, \lambda^{1/3}$ \\
  \hline
$e_{14}$ &1.361(5)  &0.657(3)   &  3.85(1)    & $3.85(7)\,\, \lambda^{1/3}$  \\
  \hline
 $e_{15}$& 1.382(7)  & 0.655(4)  &  3.92(2)   & $3.92(8)\,\, \lambda^{1/3}$ \\
 \hline
$e_{16}$&1.393(4)  & 0.657(3) & 3.97(1) & $3.97(6)\,\, \lambda^{1/3}$ \\ 
\hline
$e_{17}$ &1.457(9) & 0.663(5)    &  4.27(2)   & $4.27(4)\,\, \lambda^{1/3}$ \\
\hline
 $e_{18}$ &  1.547(7) &  0.656(4)  &   4.62(2)  & $4.62(15)\,\, \lambda^{1/3}$ \\
\hline
$e_{19}$ & 1.563(3) & 0.656(2)   &  4.70(1)   & $4.70(10)\,\, \lambda^{1/3}$ \\
\hline
$e_{20}$ &  1.567(4) & 0.663(2)  &  4.770(9)   &$4.77(7)\,\, \lambda^{1/3}$ \\
\hline
$e_{21}$& 1.572(3) & 0.665(2)   &   4.802(8)  & $4.80(7)\, \lambda^{1/3}$  \\
\hline
$e_{22}$&  1.615(4) & 0.663(2) &   5.001(9) & $5.00(12)\,\, \lambda^{1/3}$  \\
\hline
$e_{23}$ &   1.633(4)   & 0.659(2) &  5.07(1)   & $5.07(13)\,\, \lambda^{1/3}$ \\
\hline
$e_{24}$ &  1.701(5)  & 0.664(3)   &   5.46(1)  & $5.46(10)\,\, \lambda^{1/3}$  \\
\hline
$e_{25}$ &   1.716(3)  & 0.665(2)  &  5.547(9)  & $5.55(11)\,\, \lambda^{1/3}$  \\
\hline
$e_{26}$&1.728(3) & 0.666(2)    &  5.631(8)   & $5.63(6)\, \lambda^{1/3}$ \\
\hline
$e_{27}$&   1.735(2) & 0.666(1)  &  5.666(6)  & $5.67(7)\,\, \lambda^{1/3}$ \\
\hline
$e_{28}$ &  1.776(6)&  0.666(3)  &   5.90(1)  & $5.90(16)\,\, \lambda^{1/3}$ \\
\hline
$e_{29}$& 1.790(4) &  0.665(2)  &  5.98(1)  & $5.98(16)\,\, \lambda^{1/3}$ \\
\hline
 $e_{30}$ &  1.812(4) & 0.664(3  &   6.10(1)  & $6.10(20)\,\, \lambda^{1/3}$  \\
\hline
 $e_{31}$ & 1.83(1)  & 0.673(8)   &   6.32(4)  & $6.32(33)\, \lambda^{1/3}$ \\
\hline
 $e_{32}$ & 1.866(5) & 0.673(3)  &  6.52(2)   & $6.52(33)\,\, \lambda^{1/3}$ \\
\hline
 $e_{33}$ &  1.874(6)  & 0.675(4) &  6.59(2)   & $6.59(29)\,\, \lambda^{1/3}$  \\
\hline
 $e_{34}$ &  1.916(3)&  0.663(2)  & 6.76(1)    & $6.76(30)\,\, \lambda^{1/3}$ \\
\hline
 $e_{35}$ &  1.917(3) &  0.666(2)  &  6.793(9)  & $6.79(22)\,\, \lambda^{1/3}$ \\
\hline
 $e_{36}$  & 1.930(3) & 0.664(2)   & 6.87(1)   & $6.87(26)\, \lambda^{1/3}$ \\
\hline
 $e_{37}$  & 1.943(3) & 0.665(2)   &  6.960(9)   & $6.96(26)\,\, \lambda^{1/3}$  \\
\hline
\end{tabular}
\caption{Log linear fit parameters and scaling parameters $\Lambda_n$ for $n=3,...,37$ at this level of truncation. The scaling parameter in the final scaling function takes into account estimated effects of the loop truncation.}
\label{table:7}
\end{center}
\end{table}

Again we observe an excellent agreement with the expected scaling power $2/3$ of the coupling constant $g_{YM}$ for the masses of the excited states, typically with uncertainties below the $1 \% $ level\footnote{The mass $e_{12}$ is an intriguing discrepancy that may require further later re-examination. However, the quality of its fit to the scaling power law is in line with other excited states.}. The two degenerate massive lowest lying states $e_3$ and $e_4$, of particular physical relevance, are specially accurate.

\begin{figure}[h!]
    \centering
    \subfloat[Linear fit of the log of the $n=3,4,5$ masses versus $\ln g_{YM}$]{{\includegraphics[width=0.45\textwidth]{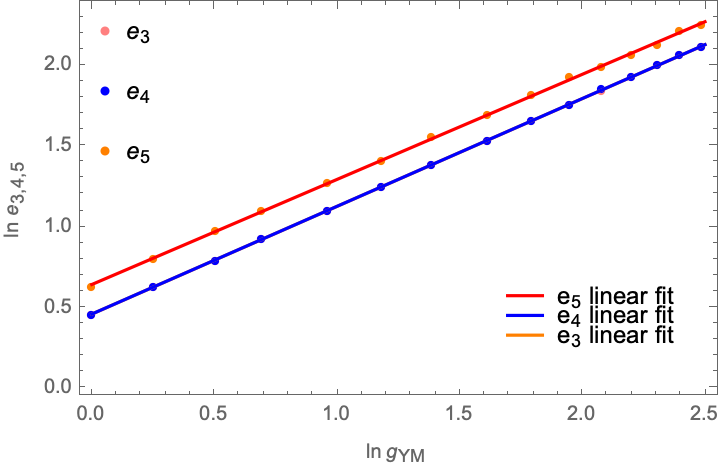} }}
    \qquad
    \subfloat[Fit of the $n=3,4,5$ masses to scaling functions $ \Lambda_{3,4,5} \, \, g_{YM}^{2/3} $ ]{{\includegraphics[width=0.45\textwidth]{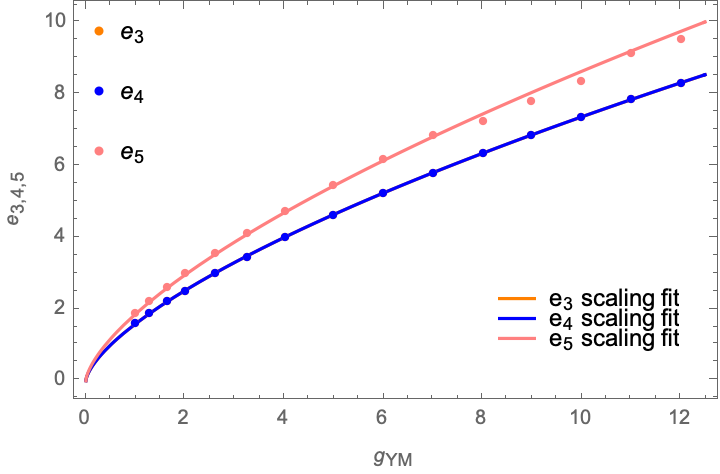} }}
    \caption{Numerical results for the masses $e_{3,4,5}\,$: logarithmic linear fits and fits to predicted scaling dependence. Note that $e_3$ is not visible due to the degeneracy with  $e_4$ }
    \label{fig:6-2PStates-vs-g_YM}
\end{figure}

\begin{figure}[h!]
    \centering
    \subfloat[Linear fit of the log of the $n=6,7,8,9$ masses  versus $\ln g_{YM}$]{{\includegraphics[width=0.45\textwidth]{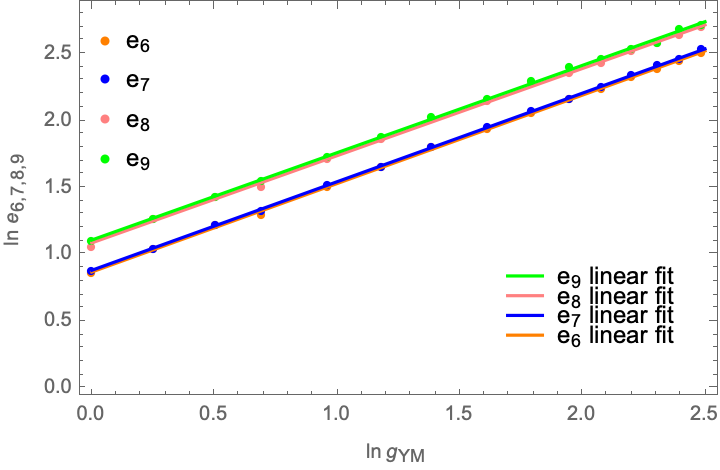} }}
    \qquad
    \subfloat[Fit of $n=6,7,8,9$ masses to scaling function $ \Lambda_{6,7,8,9} \, \, g_{YM}^{2/3} $ ]{{\includegraphics[width=0.45\textwidth]{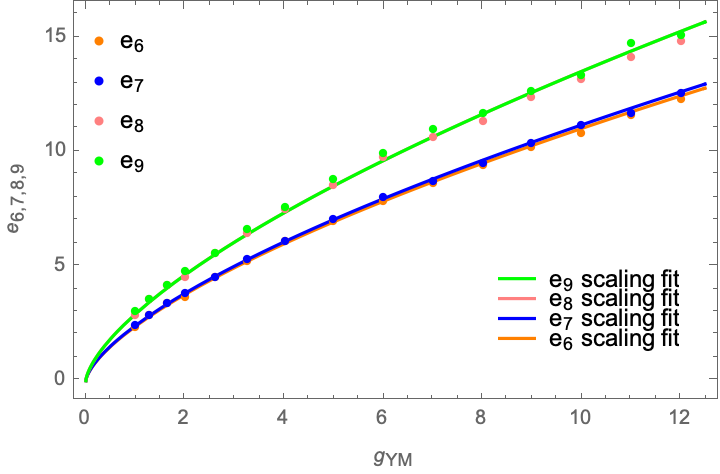} }}
    \caption{Numerical results for the masses $e_{6,7,8,9}$: logarithmic linear fits and fits to predicted scaling dependence. Two $2$-fold degenerate states are apparent.  }
    \label{fig:7-3PStates-vs-g_YM}
\end{figure}

\begin{figure}[h!]
    \centering
    \subfloat[Linear fit of the log of the $n=10,...,15$ masses  versus $\ln g_{YM}$]{{\includegraphics[width=0.45\textwidth]{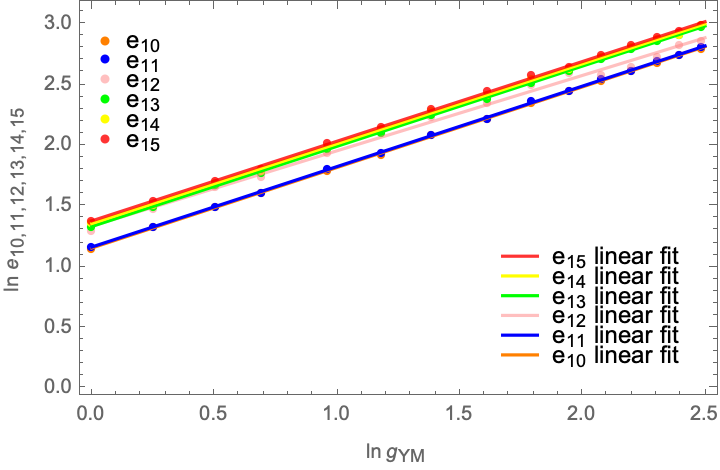} }}
    \qquad
    \subfloat[Fit of the $n=10,...,15$ masses to scaling function $ \Lambda_{10,11,12,13,14,15} \, \, g_{YM}^{2/3} $ ]{{\includegraphics[width=0.45\textwidth]{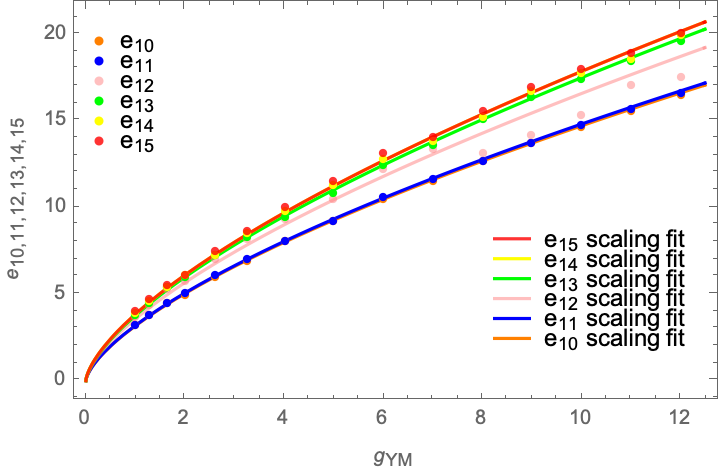} }}
    \caption{Numerical results for the masses $e_{10,11,12,13,14,15}$: logarithmic linear fits and fits to predicted scaling dependence. }
    \label{fig:8-3PStates-vs-g_YM}
\end{figure}

In figures \ref{fig:6-2PStates-vs-g_YM} , \ref{fig:7-3PStates-vs-g_YM} and \ref{fig:8-3PStates-vs-g_YM}, we display the logarithmic linear fits and the fits to the scaling power law of the numerical spectrum data for $e_n, n=3, ... , 15$. This illustrates the patterns of degeneracy, which will be further discussed in the next subsection. 

\clearpage

\subsubsection{$U(1)$ charges}

The assignment of $U(1)$ charges to the different multiplets is more cleanly carried out in a complex matrices loop basis:   

\begin{eqnarray} \label{CompReal}
[ZZ] &=& [11] + 2i [12] - [22] \nonumber \\
{[ZZ^\dagger]} &=& [11] + [22] \nonumber \\
{[Z^\dagger Z^\dagger]} &=& [11] - 2i [12] - [22] \nonumber \\
{[Z^3]} &=& [111] + 3i [112] - 3 [122] - i [222] \\
{[ZZZ^\dagger]}  &=& [111] +  i [112] + [122] + i [222] \nonumber\\
{[ZZ^\dagger Z^\dagger]} &=& [111] -  i [112] + [122] - i [222] \nonumber\\
{[Z^\dagger Z^\dagger Z^\dagger]}&=&[111] - 3i [112] - 3 [122] + i [222] \nonumber\\
&\vdots& \nonumber
\end{eqnarray}

This defines a (complex valued) linear transformation between the complex matrices loop components of a mass eigenvector and its hermitian matrices loop components. It is block diagonal, and as such one can carry out the analysis for each sector of fixed particle number.    

We find that the pattern of degeneracies is best evidenced by considering spectra corresponding to truncation subsectors within the $l_{\rm max}=14$ planar background (\cite{Koch:2021yeb}). In this subsection, we base our discussion on a $l_{\rm max}=10$ truncation for $g_{YM} = 10$. 

On general grounds, states with non-zero charges should come in doublets $\pm l$. It should be remembered, though, that the present numerical approach is based on hermitian matrices parametrized by real master variables. So typically, the eigenvector components are real. To obtain states of definite charge, the two degenerate eigenvectors have to first be expressed in a complex matrices loop component basis using (\ref{CompReal}) and then the linear combinations identified that yield two states of definite opposite 	charges. 

To illustrate the procedure, we display in table \ref{table:7.1} the $[11], [12], [22]$ components of the eigenvectors   $v^{3}, v^4$ and $v^5$. The states $e_3$ and $e_4$ are degenerate. As it can be observed from (\ref{CompReal}), they essentially "group" the real and imaginary parts of the $l=\pm 2 $ states: $v^{3}_{[11]}= - v^{3}_{[22]}$, $v^{4}_{[11]}= - v^{4}_{[22]}$  with  $v^{3,4}_{[12]}$  the imaginary part. For the singlet, one has $v^{5}_{[11]}= v^{5}_{[22]}$ with no imaginary part.

\begin{table}[h!]
\begin{center}
\begin{tabular}{||c||c|c||c||} 
\hline
& \multicolumn{2}{|c||}{Doublet} & Singlet \\
\hline
 &$v^{3}$&$v^4$ &$v^5$ \\ 
\hline
$[11]$  &0.0643&-0.0401&0.0055\\ 
 \hline
 $[12]$ &0.0830&0.1247&0.0000\\
  \hline
 $[22]$ &-0.0643& 0.0403&0.0055\\ 
\hline
\end{tabular}
\caption{The $[11], [12], [22]$ components of the eigenvectors corresponding to the states $e_3, e_4 , e_5$ }
\label{table:7.1}
\end{center}
\end{table}

After re-writing the doublet states in a complex matrices loop component basis, one performs the linear combination required to obtain the (chiral) states of definite charge. 

The choice of the linear combination coefficients is determined within the $2-$particle sector, but they act on the full eigenvector. The first $13$ components of the states are displayed in table \ref{table:7.2}. By construction, $\bar{v}^3_{[ZZ] }$ and $\bar{v}^4_{[Z^\dagger Z^\dagger] }$ are set to $1$. The table confirms the vanishing of all other components (numerically less than $1\%$), except for $\bar{v}^3_{[ZZZZ^\dagger]}$  and  $\bar{v}^4_{[ZZ^\dagger Z^\dagger Z^\dagger]}$. These $4$-particle components also carry charges $l=\pm 2$, respectively, and their enhancement to about $9\%$ results from the fact that the Hamiltonian, unlike the dilatation operator or upon radial quantization, is not number conserving. 

\begin{table}[h!]
\begin{center}
\begin{tabular}{||c||c|c||} 
\hline
 &$\bar{v}^3$& $\bar{v}^4$  \\ 
\hline
[ZZ]  &\cellcolor{blue!20} 1.000	+i0.000	&0.000	+i0.000\\ 
 \hline
 $[ZZ^\dagger]$ &-0.001	+i0.001&	-0.001- i0.001\\
 \hline
 $[Z^\dagger Z^\dagger] $ &0.000	+i0.000&	\cellcolor{blue!20} 1.000	+i0.000\\ 
 \hline
 $[ZZZ]$ &0.000	+i0.000	&0.001	+i0.002\\ 
\hline
$[ZZZ^\dagger]$ &0.000	+i0.000	&-0.005	-i0.006\\ 
\hline
$[ZZ^\dagger Z^\dagger]$ &-0.005	+i0.006	&0.000	-i0.000\\ 
\hline
$[Z^\dagger Z^\dagger Z^\dagger]$ &0.001	-i0.002	&0.000	-i0.000\\ 
 \hline
 $[ZZZZ]$ &0.001	-i0.000	&0.003	+i0.002\\ 
 \hline
 $[ZZZZ^\dagger]$ &\cellcolor{green!20} 0.091	+i0.002	&0.005	-i0.008\\ 
 \hline
 $[ZZZ^\dagger Z^\dagger]$ &0.007	+i0.002	&0.007	-i0.002\\ 
 \hline
 $[ZZ^\dagger Z Z^\dagger]$ &0.002	+i0.002	&0.002	-i0.002\\ 
 \hline
 $ZZ^\dagger Z^\dagger Z^\dagger]$ &0.005	+i0.008 &	\cellcolor{green!20} 0.091	-i0.002\\ 
 \hline
 $[Z^\dagger Z^\dagger Z^\dagger Z^\dagger]$ &0.003	-i0.002	&0.001	+i0.000\\ 
\hline
Charge & $l=2$ & $l=-2$ \\
\hline
\end{tabular}
\caption{Assignment of charges $\pm 2$ to the $3$-$4$ doublet. By construction, $\bar{v}^3_{[ZZ] }$ and $\bar{v}^4_{[Z^\dagger Z^\dagger] }$ are set to $1$.}
\label{table:7.2}
\end{center}
\end{table}
We consider now the next $4$ states $e_{6-9}$. They appear as two doubly degenerate states. Their $[111], [112], [122], [222]$ components are displayed in table \ref{table:7.3}. Again, they "encode" real and imaginary parts of definite charge states, as can be seen by inspection of (\ref{CompReal}) : $v^6_{[122]} \simeq - 3 v^6_{[111]}, v^6_{[112]} \simeq - 3 v^6_{[222]} $ and similarly for $v^7$, whereas $v^8_{[122]} \simeq v^8_{[111]}, v^8_{[112]} \simeq v^8_{[222]} $, and similarly for $v^9$.

\begin{table}[h!]
\begin{center}
\begin{tabular}{||c||c|c||c|c||} 
\hline
& \multicolumn{2}{|c||}{Doublet} &  \multicolumn{2}{|c||}{Doublet} \\
\hline
 &$v^6$&  $v^7$  &$v^8$  &$v^9$  \\ 
  \hline
 $[111]$ &0.1403&-0.1403&-0.0238&0.0174\\ 
 \hline
 $[112]$ &0.4269&0.4104&-0.0184&-0.0239\\ 
 \hline
 $[122]$ &-0.4112&0.4225&-0.0243&0.0158\\ 
 \hline
 $[222]$ &-0.1398&-0.1340&-0.0159&-0.0228\\ 
\hline
\end{tabular}
\caption{The $[111], [112], [122], [222]$ components of the eigenvectors corresponding to the states $e_6, e_7 , e_8, e_9$ }
\label{table:7.3}
\end{center}
\end{table}

Expressing the doublet states in complex matrices loop components, and performing linear combinations within each doublet to obtain states of definite charge, one obtains table \ref{table:7.4}. Other than $\bar{v}^6_{[ZZZ] }$, $\bar{v}^7_{[Z^\dagger Z^\dagger Z^\dagger] }$,$ \bar{v}^8_{[ZZZ^\dagger] }$, $\bar{v}^9_{[ZZ^\dagger Z^\dagger] } $, all set to one, the other components vanish to within $3\%$ for the $l=\pm 3$ doublet and within $5\%$ for the $l=\pm 1$ doublet. 
\begin{table}[h!]
\begin{center}
\begin{tabular}{||c||c|c||c|c||} 
\hline
& \multicolumn{2}{|c||}{Doublet} &  \multicolumn{2}{|c||}{Doublet} \\
\hline
 &$\bar{v}^6$& $\bar{v}^7$ & $\bar{v}^8$ & $\bar{v}^9$ \\ 
\hline
 [ZZ]  &0.003+i0.000&	-0.005	+i0.003&0.002	-i0.005&	-0.006	-i0.015\\ 
 \hline
 $[ZZ^\dagger]$ &-0.013	+i0.007&	-0.013	-i0.007&0.007	+i0.011&	0.007	-i0.011\\
 \hline
 $[Z^\dagger Z^\dagger] $ &-0.005	-i0.003	&0.003	-i0.000&-0.006	+i0.015	&0.002	+i0.005\\ 
 \hline
 $[ZZZ]$ &\cellcolor{blue!20} 1.000	+i0.000 &	0.000	+i0.000&0.012	-i0.013&	-0.008	+i0.001\\ 
\hline
$[ZZZ^\dagger]$ &0.011	+i0.006	&-0.003	-i0.005&\cellcolor{blue!20} 1.000	+i0.000&	0.000	+i0.000\\ 
\hline
$[ZZ^\dagger Z^\dagger]$ &-0.003	+i0.005&	0.011	-i0.006&0.000	+i0.000	&\cellcolor{blue!20} 1.000	+i0.000\\ 
\hline
$[Z^\dagger Z^\dagger Z^\dagger]$ &0.000	+i0.000&	\cellcolor{blue!20} 1.000	+i0.000&-0.008	-i0.001&	0.012	+i0.013\\ 
 \hline
 $[ZZZZ]$ &0.004	-i0.008&	-0.003	-i0.001&-0.017	+i0.060&	0.003	+i0.033\\ 
 \hline
 $[ZZZZ^\dagger]$ &-0.011	+i0.005&	0.026	-i0.026&-0.003	+i0.015&	-0.043	+i0.052\\ 
 \hline
 $[ZZZ^\dagger Z^\dagger]$ &0.026	-i0.060&	0.026	+i0.060&-0.005	-i0.026&	-0.005	+i0.026\\ 
 \hline
 $[ZZ^\dagger Z Z^\dagger]$ &0.015	-i0.009	&0.015	+i0.009&0.006	-i0.010&	0.006	+i0.010\\ 
 \hline
 $[ZZ^\dagger Z^\dagger Z^\dagger]$ &0.026	+i0.026&	-0.011	-i0.005&-0.043	-i0.052&	-0.003	-i0.015\\ 
 \hline
 $[Z^\dagger Z^\dagger Z^\dagger Z^\dagger]$ &-0.003	+i0.001&	0.004	+i0.008&0.003	-0.i033&	-0.017	-i0.060\\ 
\hline
Charge &$l=3$&  $l=-3$  &$l=1$  &$l=-1$  \\ 
\hline

\end{tabular}
\caption{Assignment of charges $\pm 3$ and  $\pm 1$ to the $6$-$9$ states.}
\label{table:7.4}
\end{center}
\end{table}

The procedure should by now be clear, and we present the assignment of non-zero charges to the states $10$-$13$ in table \ref{table:7.5}. Again, components vanish to within less than $5\%$, except for those set to $1$, and the $2$-particle components $\bar{v}^{12}_{[ZZ] }$ and $\bar{v}^{13}_{[Z^\dagger Z^\dagger] }$, also with $l=\pm 2$ charges. Interestingly, as that need not be the case, the $l=0$ states are also degenerate. A summary of the charge assignments of the states $e_{3-15}$ is presented in table \ref{table:7.6}.

\begin{table}[h!]
\begin{center}
\begin{tabular}{||c||c|c||c|c||} 
\hline
& \multicolumn{2}{|c||}{Doublet} &  \multicolumn{2}{|c||}{Doublet} \\
\hline
 &$\bar{v}^{10}$& $\bar{v}^{11}$ & $\bar{v}^{12}$ & $\bar{v}^{13}$ \\ 
\hline
 [ZZ]  &0.008+i0.004&0.005-i0.002&\cellcolor{green!20} 	-0.472+i0.007&-0.016+i0.006\\ 
 \hline
 $[ZZ^\dagger]$ &-0.005+i0.000&-0.005-i0.000&0.043+i0.020&0.0430-i0.020\\
 \hline
 $[Z^\dagger Z^\dagger] $ &0.005+i0.002&0.008-i0.004&-0.016-i0.006&\cellcolor{green!20} -0.472-i0.007\\ 
 \hline
 $[Z^3]$ &-0.004+i0.004&-0.000-i0.000&0.004-i0.018&0.009-i0.002\\ 
\hline
$[ZZZ^\dagger]$ &-0.046+i0.079&0.043-i0.004&-0.000+i0.001&-0.009+i0.012\\ 
\hline
$[ZZ^\dagger Z^\dagger]$ &0.043+i0.004&-0.046-i0.079&-0.009-i0.012&-0.000-i0.001\\ 
\hline
$[Z^\dagger Z^\dagger Z^\dagger]$ &-0.000+i0.000&-0.004-i0.004&0.009+i0.002&0.004+i0.018\\ 
 \hline
 $[ZZZZ]$ &\cellcolor{blue!20} 1.000+i0.000&-0.000+i0.000&0.021-i0.013&-0.033+i0.009\\ 
 \hline
 $[ZZZZ^\dagger]$ &-0.025-i0.002&-0.021+i0.027&\cellcolor{blue!20} 1.000+i0.000&-0.000+i0.000\\ 
 \hline
 $[ZZZ^\dagger Z^\dagger]$ &0.008+i0.005&0.008-i0.005&-0.041-i0.030&-0.041+i0.030\\ 
 \hline
 $[ZZ^\dagger Z Z^\dagger]$ &0.002+i0.001&0.002-i0.001&-0.023-i0.009&-0.023+i0.009\\ 
 \hline
 $[ZZ^\dagger Z^\dagger Z^\dagger]$ &-0.021-i0.027&-0.025+i0.002&0.000+i0.000&\cellcolor{blue!20} 1.000+i0.000\\ 
 \hline
 $[Z^\dagger Z^\dagger Z^\dagger Z^\dagger]$ &-0.000+i0.000&\cellcolor{blue!20} 1.000-i0.000&-0.033-i0.009&0.021+i0.013\\ 
\hline
Charge&$l=4$&  $l=-4$  &$l=2$  &$l=-2$  \\ 
\hline
\end{tabular}
\caption{Assignment of charges $\pm 4$ and  $\pm 2$ to the $10$-$15$ states.}
\label{table:7.5}
\end{center}
\end{table}

\begin{table}[h!]
\begin{center}
\begin{tabular}{||c|c|c||} 
\hline
 & Degeneracy & Charge  \\ 
\hline
$e_{3,4}$ & doublet &$l = \pm 2$  \\ 
  \hline
$e_{5}$ & singlet & $l=0$\\
  \hline
$e_{6,7}$& doublet &  $l=\pm 3$\\
  \hline
 $e_{8,9}$ & doublet& $l=\pm 1$ \\
  \hline
 $e_{10,11}$  &doublet& $l=\pm 4$\\
  \hline
 $e_{12,13}$ &doublet&$l=\pm 2$\\
  \hline
 $e_{14,15}$ &doublet& $l= 0$\\
\hline
\end{tabular}
\caption{Assignment of charges and degeneracies to the $3$-$15$ states..}
\label{table:7.6}
\end{center}
\end{table}

The approach can be extended for higher states, but it is clear that a more efficient and systematic way to do so would be to start directly with two complex matrices. This is under current investigation, but beyond the scope of this communication.

\subsubsection{Massless excitations}

We now turn our attention to the lowest excited sates $e_1$ and  $e_2$. Numerically, their masses do not increase with the coupling, and remain very small compared with the other massive excited states. These are the $U(N)$ traced constituent single particle states $\tr X_1$  and  $\tr X_2$, and we associate them with the non interacting (free) $U(1) \times U(1)$ subgroup of (\ref{FreeHam}). Numerically, one should be reminded that the eigenvalues of the mass matrix (\ref{eigSp}) include $N_{\rm loops}-N_{\Omega}$ unphysical zero eigenvalues, so these modes will mix with physical zero modes if present in the system. In order to confirm numerically that, indeed, our interpretation that $e_1$ and $e_2$ are decoupled zero mass states, we "switch on" masses in the Hamiltonian and seek evidence that indeed $e_1$ and $e_2$ remain decoupled states with masses equal to their "bare" masses. This will also allow us to compare our results with the few planar results available in the literature. This is carried out in the next section. 


\section{Yang-Mills coupled Hamiltonian with mass}

In this section, we consider the matrix quantum mechanical Hamiltonian 

\begin{equation}\label{MassHam}
\hat{H} = \frac{1}{2}  \sum_{A=1}^{2} \tr{P_A^2} +  \frac{m^2}{2}  \sum_{A=1}^{2} \tr{X_A^2}- \frac{g_{YM}^2}{N} \tr [X_1,X_2]^2 \, . \end{equation}

The same loop truncation with  $2615$ loops ($l_{\rm max} =14$) and $\Omega$ a $93 \times 93$ matrix is used.  We fix $m=2$ and take $g_{YM} = 0,1,...,15$.  Numerical results for the planar large $N$ energy, the planar even loop correlator values, the planar $l=0$ complex matrices loop correlator values,\footnote{The loop correlators with an odd number of any of the hermitian matrices are numerically $0$ to $4$ decimal places with very few exceptions, where their values do not exceed $3 \times 10^{-4}$, and similarly for complex matrices loop correlators with non-zero $U(1)$ charges. As such, they are not displayed.} and the first $15$ mass spectra are displayed in table \ref{table:8}.  

\begin{sidewaystable}
\begin{center}
\resizebox{1.1\textwidth}{!}{%
\begin{tabular}{||c|c|c|c|c|c|c|c|c|c|c|c|c|c|c|c|c||} 
\hline 
$g_{YM}$ & 0 &  1& 2& 3 & 4 & 5 & 6 & 7 & 8 & 9 & 10 & 11 & 12 & 13 & 14 & 15\\
\hline 
$e_{0}/N^{2}$ & 2.00000	&2.10908	&2.34396	&2.61780	&2.90090	&3.18318	&3.46097	&3.73293	&3.99883	&4.25863	&4.51261	&4.76107	&5.00439	&5.24285	&5.47664	&5.70642\\
\hline 
$\tr \, {1} /N$ &  1.0000 & 1.0000 & 1.0000 & 1.0000 & 1.0000 & 1.0000 & 1.0000 & 1.0000 & 1.0000 & 1.0000 & 1.0000 & 1.0000 & 1.0000 & 1.0000 & 1.0000 & 1.0000\tabularnewline
\hline  
$[11]\, /N^{2}$ & 0.2500&	0.2273	&0.1946&	0.1691	&0.1499	&0.1350	&0.1232	&0.1137	&0.1057	&0.0989	&0.0931	&0.0881	&0.0837	&0.0798	&0.0763	&0.0731\tabularnewline
\hline 
$[22]\, /N^{2}$ &  0.2500	&0.2273	&0.1946	&0.1691	&0.1499	&0.1350	&0.1232	&0.1136	&0.1057	&0.0989	&0.0931	&0.0881	&0.0837	&0.0798	&0.0763	&0.0731\tabularnewline
\hline
$[ZZ^{\dagger}]/N^{2}$ &0.5000 & 0.4546& 0.3892& 0.3382&0.2998&0.2700& 0.2464& 0.2273& 0.2114&0.1978& 0.1862& 0.1762& 0.1674& 0.1596& 0.1526& 0.1462\tabularnewline
\hline 
$[1111]\, /N^{3}$ &  0.1250	&0.1034	&0.0759	&0.0574	&0.0451	&0.0367	&0.0305	&0.0260	&0.0225	&0.0197	&0.0175	&0.0156	&0.0141	&0.0128	&0.0117	&0.0108\tabularnewline
\hline 
$[1122]\, /N^{3}$ & 0.0625	&0.0506	&0.0362	&0.0269	&0.0210	&0.0169	&0.0140	&0.0119	&0.0102	&0.0089	&0.0079	&0.0071	&0.0064	&0.0058	&0.0053	&0.0049\tabularnewline
\hline 
$[1212]\, /N^{3}$ &  0.0000	&0.0021	&0.0035	&0.0035	&0.0032	&0.0029	&0.0025	&0.0023	&0.0020	&0.0018	&0.0016	&0.0015	&0.0014	&0.0012	&0.0012	&0.0011\tabularnewline
\hline 
$[2222]\, /N^{3}$ & 0.1250	&0.1034	&0.0759	&0.0574	&0.0451	&0.0367	&0.0306	&0.0260	&0.0225	&0.0197	&0.0175	&0.0156	&0.0141	&0.0128	&0.0117	&0.0108\tabularnewline
\hline 
 $[ZZ^\dagger ZZ^\dagger]/N^{3}$ &0.2750&0.4050& 0.2896& 0.2154& 0.1678& 0.1352& 0.1121& 0.0950& 0.0818& 0.0714& 0.0634& 0.0566& 0.0510& 0.0464& 0.0422& 0.0390 \tabularnewline
 \hline 
$[ZZZ^\dagger Z^\dagger]/N^{3}$&0.2500& 0.2110& 0.1588& 0.1218& 0.0966& 0.0792& 0.0661& 0.0566& 0.0490&0.0430& 0.0382& 0.0342&0.031&0.0280& 0.0258& 0.0238 \tabularnewline
\hline 
$e_1$&2.0000	&2.0000	&1.9995	&1.9980	&1.9954	&1.9919	&1.9887	&1.9805	&1.9798	&1.9722	&1.9674	&1.9596	&1.9536	&1.9471	&1.9279	&1.9334 \tabularnewline
\hline
$e_2$&2.0000	&2.0000	&1.9995	&1.9981	&1.9957	&1.9925	&1.9890	&1.9819	&1.9805	&1.9728	&1.9701	&1.9606	&1.9558	&1.9483	&1.9315	&1.9384 \tabularnewline
\hline
$e_3$&4.0000	&4.2078	&4.6242	&5.0937	&5.5751	&6.0549	&6.5295	&6.9938	&7.4520	&7.8999	&8.3399	&8.7703	&9.1924	&9.6087	&10.011	&10.418 \tabularnewline
\hline
$e_4$&4.0000	&4.2080	&4.6242	&5.0937	&5.5753	&6.0556	&6.5298	&6.9946	&7.4529	&7.9006	&8.3405	&8.7714	&9.1933	&9.6094	&10.013	&10.419 \tabularnewline
\hline
$e_5$& 4.0000	&4.5327	&5.0772	&5.5688	&6.1573	&6.7943	&7.1861	&8.0037	&8.3936	&8.9599	&9.3389	&9.8451	&10.363	&10.900	&11.237	&11.649\tabularnewline
\hline
$e_6$&5.5051	&6.0541	&6.9025	&7.5558	&8.3571	&9.0595	&9.5955	&10.231	&11.104	&11.752	&12.151	&13.057	&13.534	&14.293	&14.483	&15.378 \tabularnewline
\hline
$e_7$&5.9371	&6.2794	&6.9162	&7.6186	&8.3691	&9.1177	&9.8334	&10.444	&11.232	&11.829	&12.646	&13.131	&13.799	&14.457	&14.694	&15.627 \tabularnewline
\hline
$e_8$&5.9609	&6.4296	&7.1186	&8.2597	&9.6060	&10.762	&11.366	&12.184	&13.181	&13.928	&14.740	&15.624	&15.891	&17.179	&17.700	&18.508 \tabularnewline
\hline
$e_9$&5.9697	&6.6118	&7.6290	&8.5949	&9.8060	&10.813	&11.689	&12.631	&13.481	&14.259	&14.929	&15.947	&16.264	&17.518	&18.175	&18.801 \tabularnewline
\hline
$e_{10}$&7.8321	&8.3047	&7.7023	&8.7651	&9.8374	&11.085	&11.888	&13.652	&14.408	&15.339	&15.973	&17.212	&18.158	&19.068	&19.312	&20.362 \tabularnewline
\hline
$e_{11}$&7.8697	&8.3887	&9.1365	&10.070	&11.048	&12.009	&12.887	&13.755	&14.822	&15.683	&16.549	&17.424	&18.188	&19.166	&19.669	&20.739 \tabularnewline
\hline
$e_{12}$&7.8841	&8.5108	&9.1522	&10.076	&11.100	&12.021	&12.955	&14.199	&14.831	&15.822	&16.703	&17.456	&18.381	&19.388	&19.836	&20.828 \tabularnewline
\hline
$e_{13}$& 7.9591	&8.6885	&9.8906	&11.191	&12.628	&13.785	&14.886	&15.669	&17.344	&18.062	&18.980	&20.684	&21.265	&22.673	&22.521	&24.244\tabularnewline
\hline
$e_{14}$&7.9652	&8.7147	&9.9857	&11.353	&12.656	&13.907	&15.093	&16.102	&17.492	&18.516	&19.618	&20.859	&21.667	&22.943	&22.705	&24.927 \tabularnewline
\hline
$e_{15}$&7.9911	&9.5729	&10.056	&11.655	&13.323	&14.743	&15.942	&16.575	&18.284	&19.388	&20.456	&21.426	&21.988	&23.350	&23.117	&25.383 \tabularnewline
\hline
\end{tabular}}
\caption{Numerical results obtained with a truncation to $2615$ loops ($l_{\rm max} =14$) with $\Omega$ a $93 \times 93$ matrix with $m=2$. }
\label{table:8}
\end{center}
\end{sidewaystable}

\subsection{Planar limit}

\begin{figure}[h!]
    \centering
    \subfloat[Strong coupling linear fit to $\ln (e_0 - \Lambda_0 g_{YM}^{2/3})/N^2$ versus $\ln g_{YM}$]{{\includegraphics[width=0.45\textwidth]{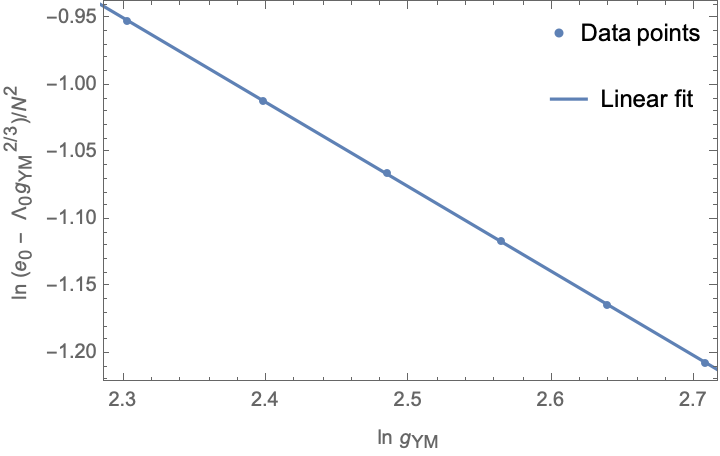} }}
    \qquad
    \subfloat[Fit of $ e_0 /N^2$ to mass corrected scaling function ]{{\includegraphics[width=0.45\textwidth]{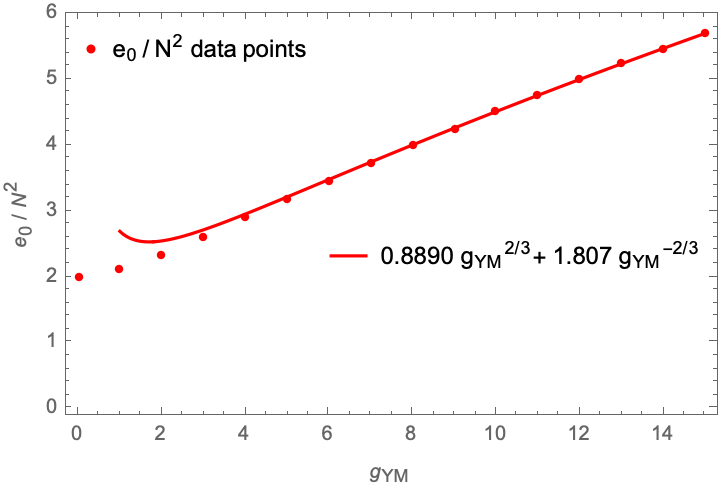} }}
    \caption{Numerically obtained large $N$ ground state energies and fit to mass corrected scaling function  }
    \label{fig:9-E0 Mass-vs-g_YM}
\end{figure}

Given that the leading large $g_{YM}$ behaviour of the large $N$ energy, that of the massless limit, has been established in the previous section, we carry out a logarithmic fit of $e_0 / N^2 - 0.8890 \,  g_{YM}^{2/3}$ to $b \, g^p_{YM}$ for large $g_{YM}$ to obtain the next, mass dependent, power dependence on $g_{YM}$. The logarithmic  fit is shown in Figure \ref{fig:9-E0 Mass-vs-g_YM} for $g_{YM} \ge 10$. The least squares fit result for the exponent is $-0.630(2)$, in other words $p=-2/3$ to a high degree of accuracy. Setting $p=-2/3$, we obtain at this truncation level:
\begin{equation*}
e_0 / N^2 = 0.8890(2) \,  \lambda^{1/3} + 1.807(4) \lambda^{-1/3} + ...
\end{equation*}

Since $m=2$, one has 
\begin{equation*}
\boxed{
e_0 / N^2 = 0.8890(2) \,  \lambda^{1/3} + 0.4518(1) \frac{m^2}{ \lambda^{1/3}} + ...}
\end{equation*}

\begin{figure}[h!]
    \centering
    \subfloat[$\tr (Z^{\dagger}Z)/N^2$  loop data and scaling asymptote]{{\includegraphics[width=0.45\textwidth]{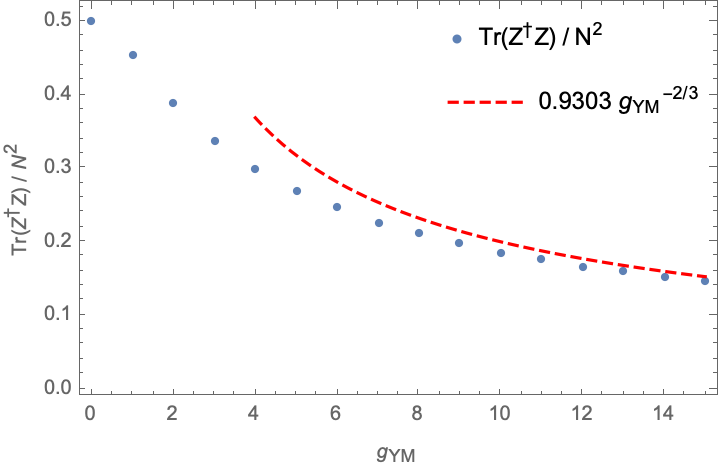} }}
    \qquad
    \subfloat[$\tr (Z^{\dagger} Z Z^{\dagger} Z) /N^3$ and $\tr (Z^{\dagger} Z^{\dagger} Z Z)/N^3$ loop data and scaling asymptotes ]{{\includegraphics[width=0.45\textwidth]{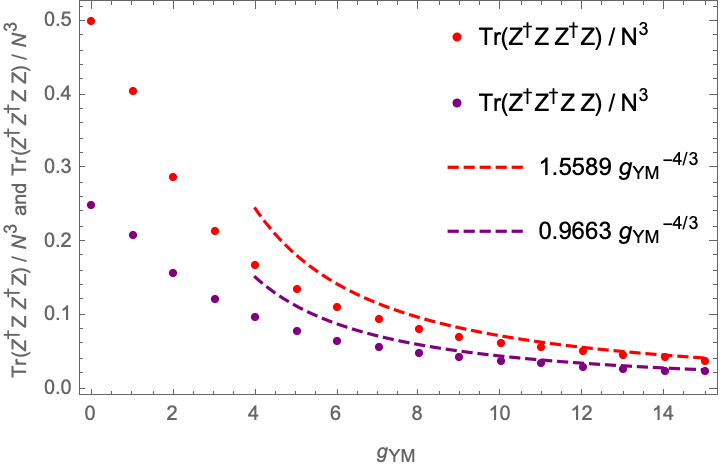} }}
    \caption{Numerically obtained large $N$ loops with two and four matrices and their scaling asymptotes  }
    \label{fig:10-ZZZZZZ Mass-vs-g_YM}
\end{figure}

For  $\ln \tr Z^{\dagger}Z/N^2$, $\tr (Z^{\dagger} Z Z^{\dagger} Z) /N^3$ and $\tr (Z^{\dagger} Z^{\dagger} Z Z)/N^3 $large $N$ correlators, we limit ourselves in displaying the large $g_{YM}$ scaling asymptotes. These are shown in Figure\ref{fig:10-ZZZZZZ Mass-vs-g_YM}

Table \ref{table:9} compares our large $N$ planar results to those available in the literature.  
\begin{table}[h!]
\begin{center}
\resizebox{1.1\textwidth}{!}
{
\begin{tabular}{||c||c|c|c||} 
\hline
& This article & \cite{Morita:2020liy} & \cite{Han:2020bkb}  \\
\hline\hline
$e_{0}/N^{2}$ & $0.8890(2)  \lambda^{1/3} + 0.4518(1) \frac{m^2}{ \lambda^{1/3}} + ...$ &$0.882\lambda^{1/3}+...$ & $0.882\lambda^{1/3} + 0.401 \frac{m^2}{ \lambda^{1/3}} + ...$ \\ 
\hline
  $\tr Z^{\dagger}Z/N^2$ & $ 0.930(1)  \lambda^{-1/3} + ... $ & $0.913 \lambda^{-1/3} + ...$  & $ 0.968  \lambda^{-1/3} + ... $   \\
\hline
\end{tabular}
}
\caption{Literature comparison}
\label{table:9}
\end{center}
\end{table}
For the large $N$ ground state energy, the improved accuracy of our approach is apparent. We also expect the leading $g_{YM}$ dependence for  $\tr Z^{\dagger}Z/N^2$ to be the most accurate, as we are able to obtain it directly in the massless limit, whereas, as is evident from Figure \ref{fig:10-ZZZZZZ Mass-vs-g_YM}, it may be difficult to extrapolate the leading dependence on $g_{YM}$ from the strong coupling regime of an Hamiltonian with mass, as is the case in \cite{Han:2020bkb}, \cite{Morita:2020liy}.

\subsection{$1/N$ spectrum}

Inspection of table \ref{table:8} shows that the energies $e_1$ and $e_2$ remain constant and very close to the "bare" mass value $m=2$\footnote{The slight decrease with $g_{YM}$ is attributed to the fact that, numerically, the unphysical "zero modes" develop small finite values. This is in opposition to the case of a quartic potential  \cite{Koch:2021yeb}, and seemingly characteristic of a Yang-Mills potential in the context of our approach}. In the massless limit then, these states remain massless, confirming the interpretation provided in the previous section. 

\begin{figure}[h!]
    \centering
    \subfloat[The planar $2$ point vertex function is:  $ - \frac{ 2 i \lambda}{m N^3} \delta_{a_1 b_2} \delta_{a_2 b_1}$ .]{{\includegraphics[width=0.45\textwidth]{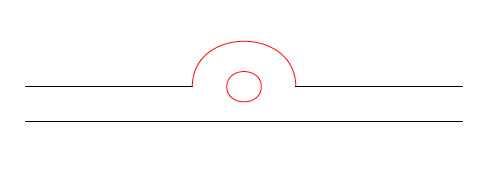} }}
    \qquad
    \subfloat[Non-planar $2$ point vertex function is $\frac{ 2 i \lambda}{m N^4} \delta_{a_1 a_2} \delta_{b_2 b_1}$.]{{\includegraphics[width=0.45\textwidth]{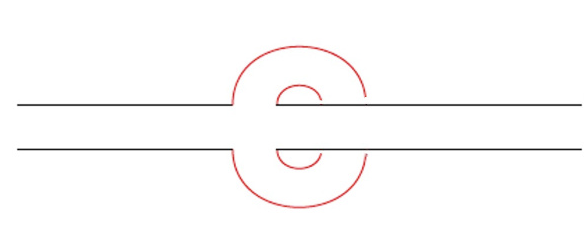} }}
    \caption{Order $\lambda=g_{YM}^2$ diagrams contributing to the $2$ point vertex function  of $<{X_1}(t_2)_{a_1 a_2}{X_1}(t_1)_{b_1 b_2}>/N^3$. Note the difference in the structure of the index contractions.}
    \label{fig:11}
\end{figure}

This may be surprising from a diagrammatic point of view. Indeed, as an example, to order $\lambda=g_{YM}^2$, one has the usual planar and non-planar contribution to the connected $2$-point function $<{X_1}(t_2)_{a_1 a_2}{X_1}(t_1)_{b_1 b_2}>/N^3$, as shown in Figure \ref{fig:11}.  

\begin{figure}[h!]
    \centering
    \subfloat[The contracted "planar" $2$ point vertex function is:  $-\frac{1}{N^2}\frac{ 2 i \lambda}{m}$ .]{{\includegraphics[width=0.45\textwidth]{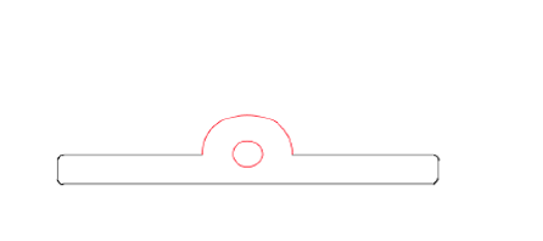} }}
    \qquad
    \subfloat[Contracted "non-planar" $2$ point vertex function: $\frac{1}{N^2}\frac{ 2 i \lambda}{m}$.]{{\includegraphics[width=0.45\textwidth]{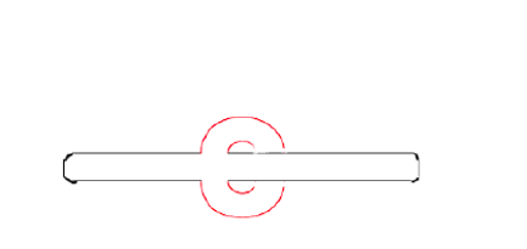} }}
    \caption{Order $\lambda=g_{YM}^2$ cancelation of contracted diagrams contributing to the $2$ point vertex function  of $<\tr {X_1}(t_2)\tr{X_1}(t_1)> /N^3$. }
    \label{fig:12}
\end{figure}

However, when the external legs are contracted, we see that the two diagrams of the previous figure cancel out, as shown in Figure \ref{fig:12}

To the same order, one can also show diagramatically, for instance, the decoupling of $1$-particle to $3$-particle states\footnote{Namely, the connected contribution to $ < \tr (X_1 X_2^2)(t_2) \tr{X_1}(t_1)> $}. 

These are, however, simple perturbative results, but numerically it is established that, non-perturbatively, the states do not receive (finite) corrections to the "bare" mass and decouple from higher (bound) states. 

Finally, for the states with energies $e_{3,4,5}$, we proceed as for $e_0/N^2$ to obtain the following mass corrected scaling functions:
\begin{align*}
e_1 &= 1.588(1) \lambda^{1/3} + 4.57(2) \lambda^{-1/3}  + ...\\
e_2 &= 1.588(1) \lambda^{1/3} + 4.56(2) \lambda^{-1/3}  + ...\\
e_3 &= 1.862(8) \lambda^{1/3} + 2.9(2) \lambda^{-1/3}  + ...
\end{align*}
or, 
\begin{align*}
e_1 &= 1.588(1) \lambda^{1/3} + 1.14(1) \frac{m^2 }{\lambda^{1/3} } + ...\\
e_2 &= 1.588(1) \lambda^{1/3} + 1.14(1) \frac{m^2}  {\lambda^{1/3} } + ...\\
e_3 &= 1.862(8) \lambda^{1/3} + 0.73(1) \frac{m^2} {\lambda^{1/3}}  + ...
\end{align*}
 
\begin{figure}[h!]
\centering
\includegraphics[width=8cm]{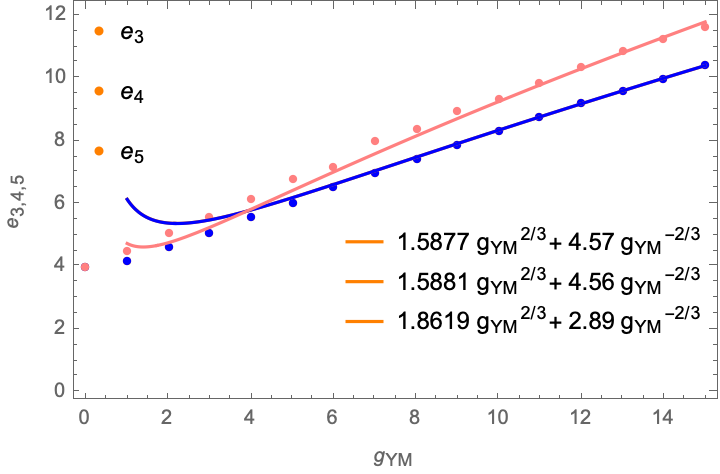}
\caption{Numerical results for the masses $e_{3,4,5}\,$ and fit to mass corrected scaling functions }
\label{fig:11-E3-5 Final-vs-g_YM}
\end{figure}

The above mass corrected scaling functions are displayed together with the numerically obtained energies in Figure \ref{fig:11-E3-5 Final-vs-g_YM}. The agreement for the two lowest lying degenerate bound states at large $g_{YM}$ is excellent.

There are no spectrum results available in the literature for comparison, as far as we know.

The assignment of charges is carried out in the same way as in the massless case, for $m=2$ and $g_{YM}=10$, and for a $l_{\rm max}=10$ mass spectrum subsector on the full $l_{\rm max}=14$ planar background. They are summarised in table \ref{table:10} \footnote{ The $e_{12}$ mass displays an irregular behaviour across the range of coupling constants, as was the case in the massless system. }. 

\begin{table}[h!]
\begin{center}
\begin{tabular}{||c|c|c||} 
\hline
n & $e_n$ & Charge  \\ 
\hline
$e_{3}$ & 8.34(1) &\multirow{2}{4em}{$l = \pm 2$}  \\ 
$e_{4}$ & 8.34(1)&   \\ 
 \hline
$e_{5}$ &9.3(4)& $l=0$\\
  \hline
$e_{6}$& 12.2(2)&  \multirow{2}{4em}{$l = \pm 3$} \\
 $e_{7}$ & 12.6(5)&\\
  \hline
 $e_{8}$& 15(1) &  \multirow{2}{4em}{$l = \pm 1$} \\
 $e_{9}$ &15(1) &\\
  \hline
$e_{10}$& 16.0(8) &  \multirow{2}{4em}{$l = \pm 4$} \\
 $e_{11}$ & 16.5(2)&\\
  \hline
$e_{12}$& 16.7(1)&  \multirow{2}{4em}{$l = \pm 2$} \\
 $e_{13}$ &19.0(2) &\\
 \hline
$e_{14}$&19.6(1) &  \multirow{2}{4em}{$l = 0$} \\
 $e_{15}$ & 20.5(9) &\\
  \hline
\end{tabular}
\caption{Assignment of charges to states $3$-$15$ for $m=2$ at $g_{YM}=10$}
\label{table:10}
\end{center}
\end{table}

\section{Discussion and outlook}
We studied the large N dynamics of two massless Yang-Mills coupled matrix quantum mechanics, by minimization of a loop truncated Jevicki-Sakita effective collective field Hamiltonian. The loop space constraints are handled by the use of master variables. 

The method was successfully applied directly in the massless limit for a range of values of the Yang-Mills coupling constant, and the scaling behaviour of different physical quantities derived from their dimensions were obtained with a high level of precision. Planar correlators of non-zero charge were shown to vanish to a high degree of accuracy, and the expected charged spectrum degeneracies confirmed for a high number of states. This attests to the validity of the method, to the consistency of the truncation scheme and to the use of master variables.  
 
We considered both planar properties of the theory, such as the large $N$ ground state energy and multi-matrix correlator expectation values, and also the $1/N$ spectrum of the theory.

For the spectrum, we established that the $U(N)$ traced fundamental constituents remain massless and decoupled from other states, and that bound states develop well defined mass gaps, with the mass of the two degenerate lowest lying bound states being determined with a particularly high degree of accuracy. These results clarify whether the system has or does not have a mass gap, and establish the nature of its patterns.

As, potentially, the presence of massless states in the spectrum may present challenges in their identification, given the nature of the numerical scheme used, we also considered the case where the fundamental constituents have a finite "bare mass", and confirmed that their $U(N)$ traced masses do not receive radiative corrections, and that they decouple from the remaining bound states in the spectrum. This also allowed us to compare some planar results with the small number of results available in the literature. As we are able to obtain the asymptotic large $\lambda$ behaviour of different physical quantities directly in the massless limit, we expect our results to be more accurate for planar quantities, and new, for the $1/N$ spectrum with finite "bare masses" as well.     

Generalization to $3$ matrices, currently under investigation, is of clear physical interest, as is the possibility of adding large $N$ quenching to the master field construction. On the "string/gravity" side, despite the absence of fermionic degrees of freedom and of a chiral limit, it is an interesting question to investigate if a BMN-type spectrum is present \cite{Berenstein:2005jq},\cite{Rodrigues:2005ec},\cite{Bhattacharyya:2008rb},\cite{Berenstein:2022srd},\cite{Pateloudis:2022oos},\cite{Lin:2022wdr}
and other dual gravity properties. In this regard, the application of the method described in this article directly to complex matrices and chiral states is of interest, and currently under investigation. The addition of fermionic degrees of freedom and, more broadly, the generalization to supersymmetric systems is, of course, of great interest. 


\section{Appendix A - Effects of truncation parameters}

In any numerical scheme involving a truncated system, it is important to try and estimate errors associated with the truncation itself, and with the different parameters of the algorithm. The accuracy of the numerical results presented in this article depend on three parameters: the truncation level  $l$ (and corresponding $l_{\rm max}$), $N$ (for a given $l$) and the convergence criterium of the optimization. We first discuss the loop truncation. 

\subsection{Loop truncation}
The results presented in the main body of this article were obtained with $l=8$, corresponding to a loop truncation involving the $2615$ single trace operators with length less or equal to $l_{\rm max}=14$, and $2652$ master variables ($N=51$). $\Omega$ is a $93 \times 93$ matrix. For a few values of $g_{YM}$ , we ran the optimization code for $l=9$, corresponding to a loop truncation involving the $8923$ single trace operators with length less or equal to $l_{\rm max}=16$, and $8930$ master variables ($N=94$). In this case, $\Omega$ is a $153 \times 153$ matrix. A number of planar quantities are displayed in tables \ref{table:B1} and \ref{table:B2} for $g_{YM}=6, 8, 10$, for both the massless case (\ref{table:B1}) and for mass $m=2$ (\ref{table:B2}). 

The first observation is how accurate ($\sim \, 0.02\%-0.04\%$) the ground state energy is with respect to truncation dependence. For loops up to quartic loops, in general the accuracy is less than $0.3\%$, except for a single loop where one can expect a truncation effect of $\sim 0.9\%$. This is incorporated in the body of the article, where we have followed a very conservative approach and in general used the least accurate estimate. 

\begin{table}[h!]
\begin{center}
\resizebox{\textwidth}{!}
{
\begin{tabular}{||c||c|c|c||c|c|c||c|c|c||} 
\hline
& \multicolumn{3}{c|}{$g_{YM}=6$} & \multicolumn{3}{c|}{$g_{YM}=8$} & \multicolumn{3}{c|}{$g_{YM}=10$}\\
\hline\hline
& $l_{\rm max}=14$ &  $l_{\rm max}=16$ & \% diff. & $l_{\rm max}=14$ & $l_{\rm max}=16$ & \% diff. &$l_{\rm max}=14$ & $l_{\rm max}=16$ & \% diff. \\
\hline
$e_{0}/N^{2}$ & 2.9355 & 2.9343 & 0.04\% &3.5562  & 3.5547 &0.04\%  & 4.1266 & 4.1249 &0.04\%  \\ 
\hline
$\tr \, {1} /N$ & 1.0000 & 1.0000  & fixed  & 1.0000 & 1.0000 & fixed  & 1.0000 & 1.0000 & fixed \\ 
\hline
$[11]\, /N^{2}$ &0.1409  & 0.1411 & -0.15 \% & 0.1163 &0.1164  & -0.17 \% &0.1002  &0.1004  &-0.16\%  \\ 
\hline
$[22]\, /N^{2}$& 0.1409 & 0.1411 & -0.16 \% & 0.1163 & 0.1164 & -0.16 \% & 0.1002 &0.1004  &-0.16\%  \\ 
\hline
$[1111]\, /N^{3}$& 0.0400 &0.0401  &-0.26\%  &0.0273  & 0.0273 &-0.28 \%  &0.0203  & 0.0203 & -0.26\% \\ 
\hline
$[1122]\, /N^{3}$ &0.0179  & 0.0179 & -0.18\% & 0.0122 & 0.0122 & -0.19\% &0.0090  &0.0091  & -0.20\% \\ 
\hline
$[1212]\, /N^{3}$ & 0.0043 & 0.0043 & -0.90\% & 0.0029 &0.0029  & -0.92\% & 0.0022 & 0.0022 & -0.98\% \\ 
 \hline
 $[2222]\, /N^{3}$ & 0.0400 & 0.0401 &-0.27\%  & 0.0273 & 0.0273 & -0.27\% &0.0202  & 0.0203 & -0.27\% \\
 \hline
 \end{tabular}
}
\caption{Loop truncation dependence of planar large $N$ energy and expectation values of single trace operators - massless case}
\label{table:B1}
\end{center}
\end{table}

\begin{table}[h!]
\begin{center}
\resizebox{\textwidth}{!}
{
\begin{tabular}{||c||c|c|c||c|c|c||c|c|c||} 
\hline
& \multicolumn{3}{c|}{$g_{YM}=6$} & \multicolumn{3}{c|}{$g_{YM}=8$} & \multicolumn{3}{c|}{$g_{YM}=10$}\\
\hline\hline
& $l_{\rm max}=14$ &  $l_{\rm max}=16$ & \% diff. & $l_{\rm max}=14$ & $l_{\rm max}=16$ & \% diff. &$l_{\rm max}=14$ & $l_{\rm max}=16$ & \% diff. \\
\hline
$e_{0}/N^{2}$ & 3.4610&	3.4604&	0.02\%   & 3.9988	&3.9979	&0.02\%   & 4.5126 &	4.5114 &0.03\%   \\ 
\hline
$\tr \, {1} /N$ & 1.0000 & 1.0000  & fixed  & 1.0000 & 1.0000 & fixed  & 1.0000 & 1.0000 & fixed \\ 
\hline
$[11]\, /N^{2}$ & 0.1232	&0.1233	&-0.07\% &  0.1057	&0.1058	&-0.10\% & 0.0931&	0.0932	&-0.11\%   \\ 
\hline
$[22]\, /N^{2}$&0.1232&	0.1233	&-0.07\%  &  0.1057&	0.1058	&-0.10\%&  0.0931	&0.0932	&-0.12\% \\ 
\hline
$[1111]\, /N^{3}$&0.0305&	0.0306	&-0.11\%   & 0.0225&	0.0225	&-0.16\%  & 0.0175 &	0.0175	&-0.18\%   \\ 
\hline
$[1122]\, /N^{3}$ & 0.0140	 &0.0140	&-0.06\% &  0.0102	&0.0102	&-0.10\%&   0.0079	&0.0079	&-0.12\%  \\ 
\hline
$[1212]\, /N^{3}$ & 0.0025&	0.0026	&-0.57\% & 0.0020&	0.0020	&-0.74\% & 0.0016&	0.0016	&-0.81\%   \\ 
 \hline
 $[2222]\, /N^{3}$ & 0.0306	&0.0306	&-0.11\% & 0.0225	&0.0225	&-0.16\% &  0.0175	&0.0175	&-0.19\%  \\
 \hline
 \end{tabular}
}
\caption{Loop truncation dependence of planar large $N$ energy and expectation values of single trace operators  with mass $m=2$}
\label{table:B2}
\end{center}
\end{table}

Tables \ref{table:B3} and \ref{table:B4} present similar results for the bound states $e_3$ to $e_{15}$. We observe the accuracy of the two degenerate lowest lying bound states ($e_{4,5}$) and their insensitivity to truncation to within ($\sim \, 0.07\%-0.13\%$).  For higher states, the effects of the truncation vary from state to state, but except for $g_{YM}=6$ in the massless case, they are found to be at most $5-6\%$, with many states much less. The case of $g_{YM}=6$ in the massless case displays larger truncation effects. Although requiring further understanding, these are the values used in the main body of the article, in line with our conservative approach to error estimation.

\begin{table}[h!]
\begin{center}
\resizebox{\textwidth}{!}
{
\begin{tabular}{||c||c|c|c||c|c|c||c|c|c||} 
\hline
& \multicolumn{3}{c|}{$g_{YM}=6$} & \multicolumn{3}{c|}{$g_{YM}=8$} & \multicolumn{3}{c|}{$g_{YM}=10$}\\
\hline\hline
& $l_{\rm max}=14$ &  $l_{\rm max}=16$ & \% diff. & $l_{\rm max}=14$ & $l_{\rm max}=16$ & \% diff. &$l_{\rm max}=14$ & $l_{\rm max}=16$ & \% diff. \\
\hline
$e_{3}$ & 5.2431	& 5.2368	&0.12\%   & 6.3508	&6.3436&	0.11\%   & 7.3699	&7.3605	&0.13\%   \\ 
\hline
$e_{4}$ &  5.2446	&5.2384	&0.12\%  & 6.3524&	6.3449 &	0.12\%  & 7.3721	&7.3621&	0.14\%  \\ 
\hline
$e_{5}$& 6.1860&	5.9766	&3.39\%   & 7.2823	&7.0897	&2.64\%   & 8.4041	&8.2057	&2.36\%   \\ 
\hline
$e_{6}$& 7.8667&	7.6625	&2.60\%  & 9.4180	&9.2940	&1.32\%   &10.863	&10.803	&0.55\%    \\ 
\hline
$e_{7}$& 8.0208&	7.8264	&2.42\%   & 9.5160 &	9.4329	&0.87\%   & 11.175&	10.861	&2.81\%   \\ 
\hline
$e_{8}$& 9.7959	&9.0238	&7.88\%  &11.338 &	11.035 &	2.66\%   & 13.163	&12.603	&4.25\%   \\ 
\hline
$e_{9}$& 9.9158	&9.2379	&6.84\%  & 11.717&	11.099&	5.27\% & 13.324	&12.969	&2.66\%   \\ 
 \hline
 $e_{10}$& 10.538	&10.437	&0.96\%  &12.661&	12.590	&0.56\%    & 14.655	&14.558	&0.66\%   \\
 \hline
 $e_{11}$&10.605&	10.466	&1.31\%   &  12.732&	12.700	&0.25\%  &14.763	&14.617	&0.99\%    \\
\hline
 $e_{12}$&  12.174	&10.946	&10.1\% & 13.181	&12.785	&3.01\%   & 15.327	&14.786	&3.53\%   \\
\hline
 $e_{13}$& 12.494&	12.303&1.53\%  &15.133	&14.682	&2.98\%  & 17.465	&17.019	&2.56\%   \\
\hline
 $e_{14}$& 12.813	&12.367&	3.49\%  & 15.233&	15.107 &	0.82\%  & 17.776	&17.460	&1.78\%   \\
  \hline
 $e_{15}$&13.202&	12.689&	3.88\%   &15.558	&15.349	&1.34\%    & 18.020	&17.683	&1.87\%   \\
\hline
 \end{tabular}
}
\caption{Loop truncation dependence of spectrum  - massless limit}
\label{table:B3}
\end{center}
\end{table}

\begin{table}[h!]
\begin{center}
\resizebox{\textwidth}{!}
{
\begin{tabular}{||c||c|c|c||c|c|c||c|c|c||} 
\hline
& \multicolumn{3}{c|}{$g_{YM}=6$} & \multicolumn{3}{c|}{$g_{YM}=8$} & \multicolumn{3}{c|}{$g_{YM}=10$}\\
\hline\hline
& $l_{\rm max}=14$ &  $l_{\rm max}=16$ & \% diff. & $l_{\rm max}=14$ & $l_{\rm max}=16$ & \% diff. &$l_{\rm max}=14$ & $l_{\rm max}=16$ & \% diff. \\
\hline
$e_{3}$ &  6.5295& 6.5252	& 0.07\%& 7.4520	&7.4459	&0.08\%     & 8.3399	&8.3317	&0.10\%  \\ 
\hline
$e_{4}$ &6.5298&	6.5255	&0.07\% & 7.4529&	7.4465	&0.09\% &8.3405 &	8.3328& 0.09\%  \\ 
\hline
$e_{5}$& 7.1861&	7.0791	&1.49\%  & 8.3936&	7.8712 &	6.22\% & 9.3389 &	8.9426& 4.24\%   \\ 
\hline
$e_{6}$& 9.5955	&9.5390	&0.59\%&  11.104	&10.704 &	3.60\%   & 12.151 &	11.991	& 1.32\%   \\ 
\hline
$e_{7}$& 9.8334	&9.5793	&2.58\% & 11.232	&10.802	&3.83\%  & 12.646	& 12.119&	4.17\%   \\ 
\hline
$e_{8}$&11.367	&10.779&	5.17\%  & 13.181	&12.046	&8.61\%  &14.740 &	13.731 &	6.85\%    \\ 
\hline
$e_{9}$& 11.689	&10.934	&6.46\%  & 13.481	&12.239 &	9.21\%   & 14.929 &	13.819&7.44\%   \\ 
 \hline
 $e_{10}$& 11.888	&11.536	&2.96\%   & 14.409	& 13.142	 &8.79\%  & 15.973 &	15.189	&4.91\%   \\
 \hline
 $e_{11}$& 12.887 &	12.985	&-0.76\%   &14.822 &	14.674	&1.00\%  & 16.549 &	16.390	& 0.96\%   \\
\hline
 $e_{12}$&12.955	&13.002	&-0.36\%   & 14.831 &	14.809& 0.15\%   & 16.703 &	16.566	& 0.82\%   \\
\hline
 $e_{13}$&14.886&	15.168	&-1.89\%   & 17.344 &	16.794	& 3.17\%  & 18.980	 & 18.739	&1.27\%   \\
\hline
 $e_{14}$&15.094 &	15.273	&-1.19\%    & 17.493	&17.405	& 0.50\%  &19.618 &	19.510	&0.55\%    \\
  \hline
 $e_{15}$& 15.942 &	15.591	&2.20\%     & 18.284 &	17.471	& 4.45\% &20.456	&19.598 &	4.19\%  \\
\hline
 \end{tabular}
}
\caption{Loop truncation dependence of spectrum   - $mass=2$}
\label{table:B4}
\end{center}
\end{table}

The parameter $gtol$ provides a convergence criterium for the optimization program, being the maximum norm of the gradient vector at the minimum.  Tables \ref{table:B5} and \ref{table:B6} compare planar and spectral data obtained with an average gradient component of $10^{-08}$ (the criterium used to obtain the results in the body of the article) versus $10^{-09}$. It is seen that they are consistently less than loop truncation effects.

\begin{table}[h!]
\begin{center}
\resizebox{0.75\textwidth}{!}
{
\begin{tabular}{||c||c|c|c||c|c|c||} 
\hline
& \multicolumn{3}{c|}{$g_{YM}=6$} & \multicolumn{3}{c|}{$g_{YM}=12$} \\
\hline\hline
${gtol}/{\sqrt{N(N+1)}}$  & $10^{-08}$ &  $10^{-09}$& \% diff.  & $10^{-08}$ &  $10^{-09}$& \% diff.   \\
\hline
$e_{0}/N^{2}$ & 2.9355 & 2.9355 & 0.00\% & 4.6600&  4.6600&0.00\% \\ 
\hline
$\tr \, {1} /N$ & 1.0000 & 1.0000  & fixed  & 1.0000 & 1.0000 & fixed  \\ 
\hline
$[11]\, /N^{2}$ &0.1409  & 0.1409 &  -0.01\% &0.0887	&0.0887	&0.00\%  \\ 
\hline
$[22]\, /N^{2}$& 0.1409 &  0.1409&  0.00\% &0.0887	&0.0887	&0.00\%   \\ 
\hline
$[1111]\, /N^{3}$& 0.0400 &0.0400 &-0.01\%  &0.0159	&0.0159	&0.00\% \\ 
\hline
$[1122]\, /N^{3}$ &0.0179  & 0.0179 & 0.00\% & 0.0071&	0.0071&	0.00\%   \\ 
\hline
$[1212]\, /N^{3}$ & 0.0043 &0.0179  & -0.02\% & 0.0017	&0.0017	&0.00\%   \\ 
 \hline
 $[2222]\, /N^{3}$ & 0.0400 & 0.0400 &-0.02\%  & 0.0159	&0.0159	&0.00\%   \\
 \hline
 \end{tabular}
}
\caption{gtol dependence of planar large $N$ energy and expectation values of single trace operators - massless case}
\label{table:B5}
\end{center}
\end{table}

\begin{table}[h!]
\begin{center}
\resizebox{0.8\textwidth}{!}
{
\begin{tabular}{||c||c|c|c||c|c|c||} 
\hline
& \multicolumn{3}{c|}{$g_{YM}=6$} & \multicolumn{3}{c|}{$g_{YM}=12$} \\
\hline\hline
${gtol}/{\sqrt{N(N+1)}}$  & $10^{-08}$ &  $10^{-09}$& \% diff.  & $10^{-08}$ &  $10^{-09}$& \% diff.   \\
\hline
$e_{3}$ & 5.2431	&	5.2444&-0.02\%   & 8.3206	&8.3203	&0.00\%  \\ 
\hline
$e_{4}$ &  5.2446	&5.2446	&0.00\%  & 8.3206	&8.3203	&0.00\%  \\ 
\hline
$e_{5}$& 6.1860&	6.1540&0.52\%   & 9.5306	 &9.5364	&-0.06\%   \\ 
\hline
$e_{6}$& 7.8667& 7.7556&1.41\%  & 12.341	&12.361	&-0.17\%      \\ 
\hline
$e_{7}$& 8.0208& 7.9610		&1.41\%   & 12.574	&12.577	&-0.03\%    \\ 
\hline
$e_{8}$& 9.7959	&	9.5779&2.23\%  &14.850	&14.891	&-0.27\%     \\ 
\hline
$e_{9}$& 9.9158	&	9.5779&1.69\%  &15.107&	15.147	&-0.27 \%  \\ 
 \hline
 $e_{10}$& 10.538	&9.7479	&1.85\%  &16.529	&16.542	&-0.07\%       \\
 \hline
 $e_{11}$&10.605&	10.533&0.68\%   &16.576	&16.579	&-0.02  \%      \\
\hline
 $e_{12}$&  12.174	&12.126	&0.39\% & 17.522	&17.543	&-0.12\%    \\
\hline
 $e_{13}$& 12.494& 12.361&1.07\%  &17.522	&17.543	&-0.12\%     \\
\hline
 $e_{14}$& 12.813	&12.574&	1.87\% &  19.990	&20.025&	-0.17\%       \\
  \hline
 $e_{15}$&13.202&12.574	&	1.26\%   &20.128	&20.151	&-0.11 \%     \\
\hline
 \end{tabular}
}
\caption{gtol dependence of spectrum  - massless limit}
\label{table:B6}
\end{center}
\end{table}

Finally, table \ref{table:B7} compares planar and spectrum results for $N=51$ ($2652$ master variables) with those obtained with $N=52$ ($2756$ master variables). Once again it is seen that the changes are consistently less than loop truncation effects.

\begin{table}[h!]
\begin{center}
\resizebox{0.5\textwidth}{!}
{
\begin{tabular}{||c||c|c|c||} 
\hline
& \multicolumn{3}{c||}{$g_{YM}=6$}  \\
\hline\hline
& $N=51$ &  $N=52$& \% diff.   \\
\hline
$e_{0}/N^{2}$ & 3.4610 & 3.4609 & 0.00\% \\ 
\hline
$\tr \, {1} /N$ & 1.0000 & 1.0000  & fixed  \\ 
\hline
$[11]\, /N^{2}$ & 0.1232	&0.1233	&-0.01\%    \\ 
\hline
$[22]\, /N^{2}$&0.1232&	0.1233	&-0.01\%  \\ 
\hline
$[1111]\, /N^{3}$ &0.0305	&0.0306	&-0.03\%  \\ 
\hline
$[1122]\, /N^{3}$ & 0.0140	&0.0140 &-0.02\%    \\ 
\hline
$[1212]\, /N^{3}$ & 0.0020&	0.0020	&-0.74\% \\ 
 \hline
 $[2222]\, /N^{3}$ &0.0025&	0.0025&	-0.13\%\\
 \hline
$e_{3}$ & 6.5295	&6.5287	&0.01\%   \\ 
\hline
$e_{4}$ & 6.5298	&6.5290	&0.01\%   \\ 
\hline
$e_{5}$& 7.1861	&7.3927	&-2.88\%  \\ 
\hline
$e_{6}$&9.5955	&9.7150	&-1.25\%      \\ 
\hline
$e_{7}$&9.8334	&9.7692	&0.65\%     \\ 
\hline
$e_{8}$& 11.367	&11.498	&-1.16\%    \\ 
\hline
$e_{9}$& 11.689	&11.740	&-0.44\% \\ 
 \hline
 $e_{10}$& 11.888	&12.700	&-6.83\%      \\
 \hline
 $e_{11}$& 12.887	&12.914	&-0.21\%   \\
\hline
 $e_{12}$& 12.955	&12.977	&-0.17\%  \\
\hline
 $e_{13}$&14.886	&15.002 &	-0.78\%    \\
\hline
 $e_{14}$& 15.094	&15.119&	-0.17\%      \\
  \hline
 $e_{15}$& 15.942	&15.855	&0.54\%   \\
\hline
 \end{tabular}
}
\caption{$N$ dependence }
\label{table:B7}
\end{center}
\end{table}

\section{Acknowledgments }
We thank Robert de Mello Koch for comments on an earlier draft of this article and Antal Jevicki for his encouragement to continue to work on loop space based numerical schemes. One of us (JPR) thanks him for his hospitality during a recent visit to the Brown Theoretical Physics Center, where the latest version of the article was completed. We also thank Xialong (Shannon) Liu for fruitful discussions.  This work is supported by the National Institute for Theoretical and Computational Sciences, NRF Grant Number 65212.

\end{document}